\pdfoutput=1

\documentclass[preprint,3p]{elsarticle}

\usepackage{lineno,hyperref}
\usepackage{bm}
\usepackage{amsmath}
\usepackage{mathtools}
\usepackage{color}
\usepackage{algorithm}
\usepackage{algorithmic}
\usepackage{multirow}
\modulolinenumbers[5]

\usepackage{amssymb}

\usepackage{subfigure}
\usepackage{array}
\newcolumntype{C}[1]{>{\centering\arraybackslash}p{#1}}

\usepackage{bigints}

\usepackage{bbding}

\usepackage{units}

\usepackage{xcolor}

\newcommand{\Div}[1]{\nabla \cdot {#1}}
\newcommand{\Grad}[1]{\nabla {#1}}
\newcommand{\Laplace}[1]{\Delta {#1}}

\usepackage{stmaryrd} % für sprungoperator

\newcommand{\avg}[1]{\{\!\{#1\}\!\}}

\newcommand{\jump}[1]{\llbracket {#1} \rrbracket }

\newcommand{\intele}[2]{ \left( {#1},{#2} \right)_{\Omega_{e}} }
\newcommand{\inteleface}[2]{ \left( {#1},{#2} \right)_{\partial\Omega_{e}} }

\newenvironment{remark}[1][Remark]{\begin{trivlist}
\item[\hskip \labelsep {\bfseries #1}]}{\end{trivlist}}

\usepackage{enumitem}
\setlist[enumerate]{label*=\roman*),ref=\roman*)}

\journal{Journal}

\begin{document}

\begin{frontmatter}

% VARIANT 1
%\title{A comparative study of compressible and incompressible\\ high-order discontinuous Galerkin Navier--Stokes solvers\\ for turbulent incompressible flows}

% VARIANT 2
%\title{On the efficiency of compressible versus incompressible\\ high-order matrix-free discontinuous Galerkin Navier--Stokes solvers\\ for turbulent incompressible flows}

% VARIANT 3
%\title{A novel high-performance matrix-free high-order\\ discontinuous Galerkin compressible Navier--Stokes solver\\ and performance comparison to incompressible solver for turbulent incompressible flows}

% VARIANT 4
%\title{A  high-order discontinuous Galerkin compressible Navier--Stokes solver\\ applied to turbulent incompressible flows and performance comparison\\ to an incompressible formulation}

% VARIANT 5
%\title{Efficiency of high-performance discontinuous Galerkin spectral element methods for under-resolved turbulent incompressible flows -- Part II: compressible versus incompressible formulations}

% VARIANT 6
%\title{A matrix-free high-order discontinuous Galerkin\\ compressible Navier--Stokes solver\\ and performance comparison for turbulent incompressible flows\\ to an incompressible formulation\medskip}

% VARIANT 7
\title{A matrix-free high-order discontinuous Galerkin\\ compressible Navier--Stokes solver:\\ A performance comparison of compressible and incompressible\\ formulations for turbulent incompressible flows}

\author{Niklas Fehn}
\ead{fehn@lnm.mw.tum.de}
\author{Wolfgang A. Wall}
\ead{wall@lnm.mw.tum.de}
\author{Martin Kronbichler\corref{correspondingauthor1}}
\cortext[correspondingauthor1]{Corresponding author at: Institute for Computational Mechanics, Technical University of Munich, Boltzmannstr. 15, 85748 Garching, Germany. Tel.: +49 89 28915300; fax: +49 89 28915301}
\ead{kronbichler@lnm.mw.tum.de}
\address{Institute for Computational Mechanics, Technical University of Munich,\\ Boltzmannstr. 15, 85748 Garching, Germany}

\begin{abstract}
Both compressible and incompressible Navier--Stokes solvers can be used and are used to solve incompressible turbulent flow problems. In the compressible case, the Mach number is then considered as a solver parameter that is set to a small value,~$\mathrm{M}\approx 0.1$, in order to mimic incompressible flows. This strategy is widely used for high-order discontinuous Galerkin discretizations of the compressible Navier--Stokes equations. The present work raises the question regarding the computational efficiency of compressible DG solvers as compared to a genuinely incompressible formulation. Our contributions to the state-of-the-art are twofold: Firstly, we present a high-performance discontinuous Galerkin solver for the compressible Navier--Stokes equations based on a highly efficient matrix-free implementation that targets modern cache-based multicore architectures. The performance results presented in this work focus on the node-level performance and our results suggest that there is great potential for further performance improvements for current state-of-the-art discontinuous Galerkin implementations of the compressible Navier--Stokes equations. Secondly, this compressible Navier--Stokes solver is put into perspective by comparing it to an incompressible DG solver that uses the same matrix-free implementation. We discuss algorithmic differences between both solution strategies and present an in-depth numerical investigation of the performance. The considered benchmark test cases are the three-dimensional Taylor--Green vortex problem as a representative of transitional flows and the turbulent channel flow problem as a representative of wall-bounded turbulent flows.
\end{abstract}


\end{frontmatter}

%\linenumbers

\section{Motivation}\label{Motivation}
Immense progress has been made in developing discontinuous Galerkin discretizations for the numerical solution of the compressible Navier--Stokes equations, see for example~\cite{Uranga2011,Chapelier2014,Beck2014b,Wiart15,Bassi2016,Fernandez2017} for recent applications to transitional and turbulent flow problems as well as references therein for details on the discretization methods. In these works, compressible DG solvers are used to also solve incompressible turbulent flow problems by using low Mach numbers. So far, a tentative conclusion might be that (explicit) compressible solvers are computationally more efficient than high-order discontinuous Galerkin formulations of the incompressible Navier--Stokes equations when applied to the solution of turbulent incompressible flow problems. However, discontinuous Galerkin discretizations of the incompressible Navier--Stokes equations that are robust in the under-resolved regime have not been available for a long time and, hence, applicability of these methods to turbulent flows has been limited. Only recently, significant progress has been made in this direction~\cite{Steinmoeller13,Joshi16,Krank2017,Fehn2017,Fehn18}. Recent results in~\cite{Fehn2018b} for the three-dimensional Taylor--Green vortex problem indicate that using an incompressible DG formulation allows to reduce the computational costs by more than one order of magnitude as compared to state-of-the-art DG implementations of the compressible Navier--Stokes equations. On the one hand, this result demonstrates that significant progress  is possible in this field, e.g, by using efficient matrix-free implementations. On the other hand, this result raises the question whether there is an intrinsic performance advantage of an incompressible formulation over well-established compressible DG solvers. This topic is highly unexplored in the DG community and requires detailed investigations. The present work extends our previous work on DG discretizations of the incompressible Navier--Stokes equations~\cite{Fehn18,Fehn2018b} towards the compressible Navier--Stokes equations and represents the first work using optimal-complexity and hardware-aware implementations for complex geometries that addresses the question regarding the computational efficiency of compressible versus incompressible high-order DG solvers when applied to the solution of under-resolved turbulent incompressible flows.

In our opinion, it is time to give particular attention to computationally efficient implementations of high-order discontinuous Galerkin discretization methods designed for and making use of modern computer hardware. While parallel scalability is often used as a motivation for DG methods~\cite{Hindenlang2012}, we recall that a primary goal in high-performance computing is optimizing the serial performance~\cite{Hager2010}. Accordingly, our focus lies on the node-level performance in the present work. In the field of computational fluid dynamics, DG methods have the potential to replace well-established discretization methods like finite volume methods and finite element methods mainly due to the fact that discontinuous Galerkin methods extend naturally to high-order of accuracy (for smooth solutions and well-resolved situations) while retaining appealing stability properties of finite volume methods~\cite{Hesthaven07}. To render high-order methods computationally efficient, a matrix-free evaluation of spatially discretized finite element operators is inevitable~\cite{Orszag1980,Kopriva09}. In the present work, we propose a high-performance discontinuous Galerkin spectral element approach for the solution of the compressible Navier--Stokes equations based on a generic matrix-free implementation with a focus on the solution of under-resolved turbulent incompressible flows. The matrix-free implementation used in this work has been shown to exhibit outstanding performance characteristics~\cite{Kronbichler2012,Kronbichler2017a,Kronbichler2017b}. High-performance implementations for high-order DG discretizations have also been proposed recently in~\cite{Muething2017}, and matrix-free implementations for continuous spectral element methods are discussed for example in~\cite{Vos2010,Cantwell2011,May2014,Kronbichler2018a}.

The present work is restricted to explicit Runge--Kutta time integration schemes~\cite{Kennedy2000,Toulorge2012,Kubatko2014} for discretizing the compressible Navier--Stokes equations in time. Explicit time integration methods for the compressible Navier--Stokes equations are attractive since they only require evaluations of the spatially discretized operator in order to advance in time without the need to solve (non-)linear systems of equations, see for example~\cite{Hindenlang2012}. For the solution of the incompressible Navier--Stokes equations, we consider splitting-type solution methods with mixed implicit/explicit treatment of viscous/convective terms for efficient time integration. Several factors have an impact on the computational efficiency of compressible solvers as compared to discretizations of the incompressible Navier--Stokes equations, the three most important ones are:
\begin{itemize}
\item Differences in time step sizes: Time step restrictions due to the CFL condition arising from an explicit formulation of the convective term are more restrictive in the compressible case than the incompressible case, i.e., the time step size is approximately inversely proportional to the Mach number~$\mathrm{M}=\frac{u}{c}$ (where~$u$ is the velocity of the fluid and~$c$ the speed of sound) in case of a compressible solver. Since a small Mach number has to be used to mimic incompressible flows, e.g.,~$\mathrm{M}\approx 0.1$, one can expect that the maximum possible time step sizes differ significantly between compressible and incompressible solvers.
\item Solution of one time step: For incompressible solvers even with an explicit treatment of the convective term, one has to at least solve linear systems of equations due to the coupling of the momentum equation and the continuity equation via the incompressibility constraint (which does not contain a time derivative term). Accordingly, the discretized finite element operators have to be applied in every solver iteration and, hence, several times within one time step. Additionally, preconditioners have to be applied (which should be realized in a matrix-free way to obtain an algorithm that is efficient for high polynomial degrees). Instead, for explicit formulations of the compressible Navier--Stokes equations it is sufficient to evaluate the operator once per time step or per Runge--Kutta stage in order to advance in time.
\item Complexity of discretized finite element operators: Discrete operators are more complex for the compressible than for the incompressible Navier--Stokes equations. The compressible Navier--Stokes equations form a system of~$d+2$ coupled equations in~$d$ space dimensions involving transformations between conserved and derived quantities as well as more complex flux computations. Nonlinearities of the equations require dealiasing strategies like over-integration that increase computational costs. In contrast, for incompressible solvers a significant amount of computational costs is spent in evaluating rather simple discrete operators such as the Laplace operator applied to a scalar quantity in the pressure Poisson equation, e.g., when using splitting methods for discretization in time.
\end{itemize}
Note that apart from these aspects other factors might have an impact on the overall performance as well, e.g., data reuse/locality and caches, parallelization and communication patterns that potentially differ for compressible and incompressible solvers (explicit vs.~implicit/iterative solvers). Comparative studies potentially suffer from the fact that the methods of investigation are not completely comparable, e.g., due to different implementation frameworks or computer hardware. Hence, we put an emphasis on using the same setup in order to provide information as valuable as possible. Both the compressible solver and the incompressible solver analyzed here make use of the same implementation framework and especially the same matrix-free kernels. Moreover, the simulations are performed on the same hardware to enable a direct comparison of the computational efficiency of both approaches. 

The contribution of the present work is twofold: i) Firstly, a high-performance implementation for the compressible Navier--Stokes equations is proposed and its computational efficiency is demonstrated for under-resolved turbulent flows with a significant gain as compared to state-of-the-art solvers; ii) Considering the differences discussed above, the question regarding which approach is more efficient appears to be unanswered. Hence, a second major concern of the present work is to perform a detailed comparison of the efficiency of compressible and incompressible high-order DG solvers for the solution of turbulent incompressible flows. We remark that performance improvements by implicit-in-time formulations are also highly relevant and should be considered as part of future work. This aspect is, however, beyond the scope of the present study.

As numerical test cases we consider the three-dimensional Taylor--Green vortex problem and the turbulent channel flow problem. We selected these test cases for two reasons. On the one hand, these benchmark problems are well-established test cases to assess transitional and turbulent flow solvers and are available in almost every CFD code due to the simple geometry of these flow problems. Hence, we expect that many groups can directly compare their results against ours without the need to invest a lot of effort into pre- or postprocessing. On the other hand, the Taylor--Green vortex problem and the turbulent channel flow problem are two of the few turbulent test cases that allow a quantitative assessment of results and are therefore well-suited for benchmarking.

The outline of the remaining of this article is as follows: In Section~\ref{CompressibleSolver}, we discuss aspects related to the temporal discretization and spatial discretization of the compressible Navier--Stokes DG solver, as well as the matrix-free implementation. We briefly summarize the incompressible Navier--Stokes DG solver in Section~\ref{IncompressibleSolver}, which is used as a reference method in the present work. In Section~\ref{EfficiencyModels}, we discuss efficiency models for the compressible solver as well as the incompressible solver and highlight differences between both formulations in terms of computational efficiency. Detailed numerical investigations demonstrating the high-performance capability of the compressible solver on the one hand and its efficiency as compared to the incompressible solver on the other hand are presented in Section~\ref{NumericalResults}. Finally, we conclude this article with a summary of our results in Section~\ref{Summary}.

\section{Compressible Navier--Stokes DG solver}\label{CompressibleSolver}

\subsection{Mathematical model}
We seek numerical solutions of the compressible Navier--Stokes equations consisting of the continuity equation, the momentum equation, and the energy equation
\begin{align}
\frac{\partial \rho}{\partial t} + \Div{\left(\rho\bm{u}\right)} &= 0\; , \label{ContinuityEquation}\\
\frac{\partial \rho\bm{u}}{\partial t} + \Div{\left(\rho \bm{u}\otimes\bm{u}\right)} + \Grad{p} -\Div{\bm{\tau}} &= \bm{f} \; ,\label{MomentumEquation}\\
\frac{\partial \rho E}{\partial t} + \Div{\left(\rho E \bm{u}\right)} + \Div{\left(p \bm{u}\right)} -\Div{\left(\bm{\tau}\cdot\bm{u}\right)} + \Div{\bm{q}} &= \bm{f}\cdot\bm{u}\; , \label{EnergyEquation}
\end{align}
where~$\bm{f}$ denotes the body force vector,~$\rho$ the density of the fluid,~$\bm{u}$ the velocity vector,~$p$ the pressure,~$E$ the total energy per unit mass,~$\bm{\tau}$ the viscous stress tensor, and~$\bm{q}$ the heat flux. We assume a Newtonian fluid with zero bulk viscosity
\begin{align}
\bm{\tau}(\Grad{\bm{u}}) = \mu \left(\Grad{\bm{u}}+\left(\Grad{\bm{u}}\right)^T - \frac{2}{3}\left(\Div{\bm{u}}\right)\bm{I}\right) \; ,
\end{align}
and assume that the heat flux vector~$\bm{q}$ can by described by Fourier's law of heat conduction
\begin{align}
\bm{q}( \Grad{T}) = - \lambda \Grad{T} \; ,
\end{align}
where~$\lambda$ is the thermal conductivity. The above~$d+2$ equations contain~$d+4$ unknowns ($\rho,\bm{u},E,p,T$), where~$d$ denotes the number of space dimensions. Hence, two additional equations are necessary to close the system of equations. These equations are the equation of state and an equation for the specific energy. We assume an ideal gas law,~$p = \rho R T$, as well as a calorically perfect gas,~$E = e + \bm{u}^{2} / 2=c_{v} T + \bm{u}^{2} / 2$, where~$c_{v} = R/(\gamma-1)$ is the specific heat at constant volume with the specific gas constant~$R$ and the ratio of specific heats~$\gamma=c_p/c_v$. Using the vector~$\bm{U}$ of conserved variables
\begin{align}
\bm{U} = \begin{pmatrix}
\rho \\
\rho \bm{u} \\
\rho E
\end{pmatrix}  \in \mathbb{R}^{d+2} \; ,
\end{align}
the compressible Navier--Stokes equations can be written in a compact form as
\begin{align}
\frac{\partial \bm{U}}{\partial t} + \Div{\bm{F}_{\mathrm{c}}\left(\bm{U}\right)} - \Div{\bm{F}_{\mathrm{v}}\left(\bm{U},\Grad{\bm{U}}\right)} = \bm{F}_{\mathrm{rhs}} \; ,\label{CompressibleNavierStokesCompact}
\end{align}
where the convective flux, the viscous flux, and the right-hand side vector are
\begin{align}
\bm{F}_{\mathrm{c}}\left(\bm{U}\right) = 
\begin{pmatrix}
\left(\rho \bm{u}\right)^T\\
\rho \bm{u}\otimes\bm{u} + p \bm{I}\\
\left(\rho E + p\right)\bm{u}^T
\end{pmatrix} \; , \;
\bm{F}_{\mathrm{v}}\left(\bm{U},\Grad{\bm{U}}\right) = 
\begin{pmatrix}
0\\
\bm{\tau}\left(\Grad{\bm{u}}\right)\\
\left(\bm{\tau}\left(\Grad{\bm{u}}\right)\cdot \bm{u}\right)^T - \bm{q}\left(\Grad{T}\right)^T
\end{pmatrix} \; , \;
\bm{F}_{\mathrm{rhs}}=
\begin{pmatrix}
0\\
\bm{f}\\
\bm{f}\cdot\bm{u}
\end{pmatrix} \; . \label{CompressibleNavierStokesFluxes}
\end{align}
The following derivations of the weak discontinuous Galerkin formulation make use of the fact that the viscous flux depends linearly on the gradient of the conserved variables~$\bm{U}$, i.e.,
\begin{align}
\bm{F}_{\mathrm{v}}\left(\bm{U},\Grad{\bm{U}}\right) = \bm{G}\left(\bm{U}\right) : \Grad{\bm{U}} \; ,
\end{align}
using a tensor~$\bm{G}$ as introduced in~\cite{Hartmann2005}.

\subsection{Discontinuous Galerkin formulation}
For discretization in space we use a high-order discontinuous Galerkin approach, see for example~\cite{Hesthaven07}. The Lax--Friedrichs flux is used to discretize the convective term and the symmetric interior penalty method for the viscous term. For reasons of brevity, a detailed description of the imposition of boundary conditions is omitted since we mainly consider problems with periodic boundary conditions in this work and since we want to focus on algorithmic aspects of the Navier--Stokes solver.

\subsubsection{Notation}
The computational domain~$\Omega_h = \bigcup_{e=1}^{N_{\text{el}}} \Omega_{e}$ with boundary~$\Gamma_h = \partial \Omega_h$ approximating the physical domain~$\Omega \subset \mathbb{R}^d$ is composed of~$N_{\text{el}}$ non-overlapping quadrilateral/hexahedral elements. When using discontinuous Galerkin discretizations, the solution is approximated by polynomials inside each element. We define the space~$\mathcal{V}_{h}$ of test functions~$\bm{V}_h\in \mathcal{V}_{h}$ and trial functions~$\bm{U}_h\in \mathcal{V}_{h}$ as
\begin{align}
\mathcal{V}_{h} &= \left\lbrace\bm{U}_h\in \left[L_2(\Omega_h)\right]^{d+2}\; : \; \bm{U}_h\left(\bm{x}(\boldsymbol{\xi})\right)\vert_{\Omega_{e}}= \tilde{\bm{U}}_h^e(\boldsymbol{\xi})\vert_{\tilde{\Omega}_{e}}\in \mathcal{V}_{h}^{e} = [\mathcal{P}_{k}(\tilde{\Omega}_{e})]^{d+2}\; ,\;\; \forall e=1,\ldots,N_{\text{el}} \right\rbrace\; .
\end{align}
Here,~$\tilde{\Omega}_e=[0,1]^d$ is the reference element with reference coordinates~$\boldsymbol{\xi}$. The mapping from reference space to physical space is denoted by~$\bm{x}^e(\boldsymbol{\xi}) : \tilde{\Omega}_e \rightarrow \Omega_e$, and~$\mathcal{P}_{k}(\tilde{\Omega}_{e})$ is the space of polynomials of tensor degree~$\leq k$. To construct multidimensional shape functions as the tensor product of one-dimensional shape functions, we use a nodal basis with Lagrange polynomials based on the Legendre--Gauss--Lobatto (LGL) support points. We also introduce the space
\begin{align}
\mathcal{R}_{h} &= \left\lbrace\bm{W}_h\in \left[L_2(\Omega_h)\right]^{(d+2)\times d}\; : \; \bm{W}_h\left(\bm{x}(\boldsymbol{\xi})\right)\vert_{\Omega_{e}}= \tilde{\bm{W}}_h^e(\boldsymbol{\xi})\vert_{\tilde{\Omega}_{e}}\in [\mathcal{P}_{k}(\tilde{\Omega}_{e})]^{(d+2)\times d}\; ,\;\; \forall e=1,\ldots,N_{\text{el}} \right\rbrace
\end{align}
for the auxiliary variable~$\bm{Q}_h\in \mathcal{R}_{h}$ and the corresponding weighting functions~$\bm{R}_h\in \mathcal{R}_{h}$ that are needed in the following when deriving the discontinuous Galerkin formulation. We use an element-based notation of the weak formulation using the common abbreviations~$\intele{v}{u} = \int_{\Omega_e} v \odot u \; \mathrm{d}\Omega$ for volume integrals and~$\inteleface{v}{u} = \int_{\partial \Omega_e} v \odot u \; \mathrm{d} \Gamma$ for surface integrals. To define numerical fluxes on a face~$f=\partial \Omega_{e^-} \cap \partial \Omega_{e^+}$ between two adjacent elements~$e^-$ and~$e^+$ we introduce the average operator~$\avg{u} = (u^- + u^+)/2$  and the jump operator~$ \jump{u} = u^- \otimes \bm{n}^- + u^+ \otimes \bm{n}^+$. The generic operator~$\odot$ symbolizes inner products, e.g.~$\bm{v}\cdot \bm{u}$ for rank-1 tensors and~$\bm{v}:\bm{u}$ for rank-2 tensors, while~$\otimes$ denotes tensor products. Moreover,~$(\cdot)^-$ denotes interior information,~$(\cdot)^+$ exterior information from the neighboring element, and~$\bm{n}$ the outward pointing unit normal vector.

\subsubsection{Derivation of weak DG formulation}
The derivation of the weak formulation and especially the symmetric interior penalty method used to discretize the viscous term follows~\cite{Hartmann2005}. In a first step, the system of equations containing second derivatives is written as a system of first order equations, see also~\cite{Hesthaven07}. By introducing the auxiliary variable~$\bm{Q}=\bm{F}_{\mathrm{v}}\left(\bm{U},\Grad{\bm{U}}\right)$ we obtain
\begin{align}
\frac{\partial \bm{U}}{\partial t} + \Div{\bm{F}_{\mathrm{c}}\left(\bm{U}\right)} - \Div{\bm{Q}} = \bm{F}_{\mathrm{rhs}} \; ,\\
\bm{Q} = \bm{F}_{\mathrm{v}}\left(\bm{U},\Grad{\bm{U}}\right)\; .
\end{align}
In a second step the equations are multiplied by weighting functions~$\bm{V}_h \in \mathcal{V}_{h}$ and~$\bm{R}_h \in \mathcal{R}_{h}$ and integrated over~$\Omega_h$. Subsequently, integration by parts is performed and numerical fluxes~$\left(\cdot\right)^{*}$ are introduced to obtain
\begin{align}
\begin{split}
\intele{\bm{V}_h}{\frac{\partial\bm{U}_h}{\partial t}} 
- \intele{\Grad{\bm{V}_h}}{\bm{F}_{\mathrm{c}}\left(\bm{U}_h\right)} + \inteleface{\bm{V}_h}{\bm{F}_{\mathrm{c}}^*\left(\bm{U}_h\right)\cdot \bm{n}}\\
+ \intele{\Grad{\bm{V}_h}}{\bm{Q}_h} - \inteleface{\bm{V}_h}{\bm{Q}_h^* \cdot \bm{n}} = \intele{\bm{V}_h}{\bm{F}_{\mathrm{rhs}}} \; ,
\end{split}\label{WeakFormEquation1}
\end{align}
and for the auxiliary variable
\begin{align}
\begin{split}
\intele{\bm{R}_h}{\bm{Q}_h} &= \intele{\bm{R}_h}{\bm{F}_{\mathrm{v}}\left(\bm{U}_h,\Grad{\bm{U}_h}\right)} = \intele{\bm{R}_h:\bm{G}\left(\bm{U}_h\right)}{\Grad{\bm{U}_h}} \\
&= -\intele{\Div{\left(\bm{R}_h:\bm{G}\left(\bm{U}_h\right)\right)}}{\bm{U}_h} + \inteleface{\bm{R}_h:\bm{G}\left(\bm{U}_h\right)}{\bm{U}_h^* \otimes \bm{n}} \\
&= \intele{\bm{R}_h:\bm{G}\left(\bm{U}_h\right)}{\Grad{\bm{U}_h}} + \inteleface{\bm{R}_h:\bm{G}\left(\bm{U}_h\right)}{\left(\bm{U}_h^*-\bm{U}_h\right) \otimes \bm{n}}\\
&= \intele{\bm{R}_h}{\bm{F}_{\mathrm{v}}\left(\bm{U}_h,\Grad{\bm{U}_h}\right)} + \inteleface{\bm{R}_h}{\bm{F}_{\mathrm{v}}\left(\bm{U}_h,\left(\left(\bm{U}_h^*-\bm{U}_h\right) \otimes \bm{n}\right)\right)}\; .
\end{split}\label{WeakFormEquation2}
\end{align}
As usual when deriving the interior penalty formulation, the above equation with the auxiliary variable is integrated by parts twice, where the numerical flux is used in the first integration by parts step and the interior flux is used in the second integration by parts step. Choosing~$\bm{R}_h = \Grad{\bm{V}_h}$, equation~\eqref{WeakFormEquation2} can be inserted into equation~\eqref{WeakFormEquation1} to obtain the primal formulation: Find~$\bm{U}_h\in \mathcal{V}_h$ such that
\begin{align}
\begin{split}
\intele{\bm{V}_h}{\frac{\partial\bm{U}_h}{\partial t}} 
- \intele{\Grad{\bm{V}_h}}{\bm{F}_{\mathrm{c}}\left(\bm{U}_h\right)} + \inteleface{\bm{V}_h}{\bm{F}_{\mathrm{c}}^*\left(\bm{U}_h\right)\cdot \bm{n}}
+ \intele{\Grad{\bm{V}_h}}{\bm{F}_{\mathrm{v}}\left(\bm{U}_h,\Grad{\bm{U}_h}\right)}\\ - \inteleface{\bm{V}_h}{\bm{Q}_h^* \cdot \bm{n}} + \inteleface{\Grad{\bm{V}_h}}{\bm{F}_{\mathrm{v}}\left(\bm{U}_h,\left[\left(\bm{U}_h^*-\bm{U}_h\right) \otimes \bm{n}\right]\right)} = \intele{\bm{V}_h}{\bm{F}_{\mathrm{rhs}}} \; ,
\end{split}\label{PrimalFormulation}
\end{align}
for all~$\bm{V}_h\in \mathcal{V}_h^{e}$ and for all elements~$e=1,...,N_{\mathrm{el}}$. We use the local Lax--Friedrichs flux for the convective term~\cite{Hesthaven07}
\begin{align}
\bm{F}_{\mathrm{c}}^*\left(\bm{U}_h\right) = \avg{\bm{F}_{\mathrm{c}}\left(\bm{U}_h\right)} + \frac{\Lambda}{2}\jump{\bm{U}_h} \; ,
\end{align}
where~$\Lambda= \max\left(\Vert \bm{u}_h^-\Vert + c_h^-,\Vert \bm{u}_h^+\Vert + c_h^+\right)$ with  the speed of sound~$c_h=\sqrt{\vert\gamma R T_h\vert}=\sqrt{\vert\gamma p_h/\rho_h\vert}$. For the viscous term, the interior penalty method~\cite{Arnold2002} is used
\begin{align}
\bm{Q}_h^* = \avg{\bm{F}_{\mathrm{v}}\left(\bm{U}_h,\Grad{\bm{U}_h}\right)}-\bm{\tau}_{\mathrm{IP}} \jump{\bm{U}_h} \; , \\
\bm{U}_h^* = \avg{\bm{U}_h} \; .
\end{align}
As in~\cite{Hartmann2005}, the diagonal penalization matrix~$\bm{\tau}_{\mathrm{IP}}$ is chosen as a scaled identity matrix,~$\bm{\tau}_{\mathrm{IP}}=\tau_{\mathrm{IP}}\bm{I}$, with a scalar penalty parameter~$\tau_{\mathrm{IP}}$ using the definition according to~\cite{Hillewaert13}
\begin{align}
\tau_{\mathrm{IP},e} = \nu (k+1)^2 \frac{A\left(\partial \Omega_e \setminus \Gamma_h\right)/2 + A\left(\partial \Omega_e \cap \Gamma_h\right)}{V\left(\Omega_e\right)}\; ,\label{TauIP_Element}
\end{align}
where~$V\left(\Omega_e\right) = \int_{\Omega_e}\mathrm{d}\Omega$ is the element volume and~$A(f) = \int_{f\subset\partial\Omega_e}\mathrm{d}\Gamma$ the surface area. On interior faces, the penalty parameter~$\tau_{\mathrm{IP}}$ is obtained by taking the maximum value of both elements adjacent to face~$f$. Note that we multiply the penalty factor by the kinematic viscosity~$\nu=\mu/\rho_0$ with a reference density~$\rho_0$ to obtain consistent physical units. By inserting the numerical fluxes into the primal formulation~\eqref{PrimalFormulation} we arrive at the following weak discontinuous Galerkin formulation: Find~$\bm{U}_h\in \mathcal{V}_h$ such that
\begin{align}
\begin{split}
\intele{\bm{V}_h}{\frac{\partial\bm{U}_h}{\partial t}} 
- \intele{\Grad{\bm{V}_h}}{\bm{F}_{\mathrm{c}}\left(\bm{U}_h\right)} + \inteleface{\bm{V}_h}{\left(\avg{\bm{F}_{\mathrm{c}}\left(\bm{U}_h\right)} + \frac{\Lambda}{2}\jump{\bm{U}_h}\right)\cdot \bm{n}}
+ \intele{\Grad{\bm{V}_h}}{\bm{F}_{\mathrm{v}}\left(\bm{U}_h,\Grad{\bm{U}_h}\right)} \\
- \inteleface{\bm{V}_h}{\left(\avg{\bm{F}_{\mathrm{v}}\left(\bm{U}_h,\Grad{\bm{U}_h}\right)}-\tau_{\mathrm{IP}} \jump{\bm{U}_h}\right) \cdot \bm{n}}
-\frac{1}{2} \inteleface{\Grad{\bm{V}_h}}{\bm{F}_{\mathrm{v}}\left(\bm{U}_h,\jump{\bm{U}_h}\right)} = \intele{\bm{V}_h}{\bm{F}_{\mathrm{rhs}}} \; ,
\end{split}\label{WeakFormulationConservedVariables}
\end{align}
for all~$\bm{V}_h\in \mathcal{V}_h^{e}$ and for all elements~$e=1,...,N_{\mathrm{el}}$.

\subsubsection{Numerical calculation of integrals} 
To implement the above discontinuous Galerkin method, the volume and surface integrals occurring in the weak formulation~\eqref{WeakFormulationConservedVariables} have to be evaluated numerically. For these derivations we restrict our presentation to the terms involved in the Euler equations for reasons of simplicity, i.e., the viscous term is neglected,
\begin{align}
\int_{\Omega_e}\bm{V}_h\cdot\frac{\partial\bm{U}_h}{\partial t}\mathrm{d}\Omega
- \int_{\Omega_e}\Grad{\bm{V}_h} : \bm{F}_{\mathrm{c}}\left(\bm{U}_h\right)\mathrm{d}\Omega  + \int_{\partial\Omega_e} \bm{V}_h \cdot \left(\bm{F}^*_{\mathrm{c}}\left(\bm{U}_h\right)\cdot \bm{n}\right) \mathrm{d}\Gamma
= \int_{\Omega_e}\bm{V}_h \cdot \bm{F}_{\mathrm{rhs}} \mathrm{d}\Omega \; .\label{IntegralsPhysicalSpace}
\end{align} 
In a first step, the integrals are transformed from pyhsical space~$\bm{x}$ to reference space~$\bm{\xi}$
\begin{align}
\begin{split}
\int_{\tilde{\Omega}_e}\bm{V}_h(\bm{\xi})\cdot\frac{\partial\bm{U}_h(\bm{\xi})}{\partial t}\vert \det \bm{J}^e(\bm{\xi}) \vert\mathrm{d}\bm{\xi}
- \int_{\tilde{\Omega}_e}\left(\nabla_{\bm{\xi}}\bm{V}_h(\bm{\xi})\cdot (\bm{J}^e)^{-1}(\bm{\xi})\right) : \bm{F}_{\mathrm{c}}\left(\bm{U}_h(\bm{\xi})\right)\vert \det \bm{J}^e(\bm{\xi}) \vert\mathrm{d}\bm{\xi}\\
 + \int_{\partial\tilde{\Omega}_e} \bm{V}_h(\bm{\xi}) \cdot \left(\bm{F}^*_{\mathrm{c}}(\bm{U}_h(\bm{\xi}))\cdot \bm{n}(\bm{\xi})\right) \Vert \det (\bm{J}^e(\bm{\xi})) (\bm{J}^e)^{-T}(\bm{\xi})\cdot \bm{n}_{\bm{\xi}}\Vert\mathrm{d}\bm{\xi}_f\\
= \int_{\tilde{\Omega}_e}\bm{V}_h(\bm{\xi}) \cdot \bm{F}_{\mathrm{rhs}}(\bm{\xi}) \vert \det \bm{J}^e(\bm{\xi}) \vert\mathrm{d}\bm{\xi} \; ,
\end{split}\label{IntegralsReferenceSpace}
\end{align}
where~$\bm{J}^e = \partial \bm{x}^e/\partial \bm{\xi}$ is the Jacobian matrix of the mapping~$\bm{x}^e(\bm{\xi})$ and~$\bm{n}_{\bm{\xi}}$ the normal vector in reference space~$\bm{\xi}$.
For the numerical calculation of integrals in reference space we apply Gaussian quadrature ensuring exact integration of polynomials up to degree~$2 n_q - 1$ when using~$n_q$ Gauss--Legendre quadrature points in one space dimension. The integrals in equation~\eqref{IntegralsReferenceSpace} can be written as a sum over all quadrature points, which gives
\begin{align}
\begin{split}
\sum_{q=1}^{N_{q,e}}\bm{V}_h(\bm{\xi}_q)\cdot\frac{\partial\bm{U}_h(\bm{\xi}_q)}{\partial t}\vert \det \bm{J}^e(\bm{\xi}_q) \vert w_q
- \sum_{q=1}^{N_{q,e}}\left(\nabla_{\bm{\xi}}\bm{V}_h(\bm{\xi}_q)\cdot (\bm{J}^e)^{-1}(\bm{\xi}_q)\right) : \bm{F}_{\mathrm{c}}\left(\bm{U}_h(\bm{\xi}_q)\right)\vert \det \bm{J}^e(\bm{\xi}_q) \vert w_q\\
 + \sum_{f=1}^{N_{\mathrm{faces}}}\sum_{q=1}^{N_{q,f}} \bm{V}_h(\bm{\xi}_q) \cdot \left(\bm{F}^*_{\mathrm{c}}(\bm{U}_h(\bm{\xi}_q))\cdot \bm{n}(\bm{\xi}_q)\right) \Vert \det (\bm{J}^e(\bm{\xi}_q)) (\bm{J}^e)^{-T}(\bm{\xi}_q)\cdot \bm{n}_{\bm{\xi}}\Vert w_q\\
= \sum_{q=1}^{N_{q,e}}\bm{V}_h(\bm{\xi}_q) \cdot \bm{F}_{\mathrm{rhs}}(\bm{\xi}_q) \vert \det \bm{J}^e(\bm{\xi}_q) \vert w_q \; .
\end{split}\label{IntegralsNumericalQuadrature}
\end{align}
For the calculation of the mass matrix term and the body force term we use~$n_q=k+1$ quadrature points. The mass matrix term contains polynomials of degree~$2k$ as long as the geometry terms are constant. While this term is integrated exactly on affine element geometries, a quadrature error due to geometrical nonlinearities is introduced on more complex element geometries. However, the geometry is typically well-resolved so that geometrical nonlinearities are typically less of a concern as compared to nonlinearities arising from the equations such as the nonlinearity of the convective term and the viscous term, see also~\cite{Mengaldo2015}. A major concern for compressible spectral element solvers is therefore the selection of the number of quadrature points for the integration of the convective and viscous terms, see for example~\cite{Beck2014b,Gassner2013} in the context of discontinuous Galerkin discretizations of the compressible Navier--Stokes equations,~\cite{Mengaldo2015,Kirby2003} in the context of continuous spectral element methods, and~\cite{Hesthaven07} for a more general discussion of the concepts. Since these terms contain nonlinearities, a standard quadrature with~$k+1$ quadrature points introduces quadrature errors, also known as aliasing, that might lead to instabilities especially for under-resolved situations. While techniques such as filtering~\cite{Gassner2013,Fischer2001,Flad2016} and split form DG methods~\cite{Flad2017,Winters2017} can be used as dealiasing strategies, we entirely focus on consistent integration (also known as over-integration or polynomial dealiasing) in the present work as a mechanism to avoid aliasing effects, see for example~\cite{Beck2014b,Gassner2013}. Using this strategy, the number of quadrature points is increased,~$n_q > k+1$, in order to eliminate or minimize quadrature errors. This approach is straight forward for the incompressible Navier--Stokes equations since the nonlinear terms are still polynomials (of higher degree) in this case. For the compressible Navier--Stokes equations, however, the nonlinear terms involve rational functions (one has to divide by the density to obtain derived quantities such as the temperature or the pressure) rendering an exact integration with a finite number of quadrature points difficult. Accordingly, a pragmatic approach is in order to keep the computational costs for the evaluation of integrals feasible. Assuming that the density is constant,~$\rho = \mathrm{const}$, one can show that the compressible Navier--Stokes equations contain at most cubic nonlinearities of degree~$3k$, so that~$n_q=\lceil \frac{4k+1}{2}\rceil = 2k+1$ quadrature points would be sufficient for exact integration. However, numerical results~\cite{Beck2014b} give evidence that a 3/2-dealiasing strategy is already sufficient for problems where the density variations are small such as the incompressible flow problems considered in the present work. Hence, we apply this latter strategy in the present work and use~$n_q = \lceil \frac{3k+1}{2} \rceil$ quadrature points (3/2-rule) that would be necessary to exactly integrate quadratic nonlinearities of degree~$2k$. However, to demonstrate the computational efficiency and the additional costs due to over-integration we will analyze the performance of the present matrix-free implementation for a standard quadrature rule with~$k+1$ points as well as over-integration strategies with~$\lceil \frac{3k+1}{2} \rceil$ and~$2k+1$ points. Let us remark that despite consistent over-integration, instabilities might still occur in some cases, see for example~\cite{Winters2017,Moura2017} in the context of the inviscid Taylor--Green vortex problem. Since we also use a 3/2-dealiasing strategy for the integration of the nonlinear terms of the incompressible solver, this setup allows a fair comparison of the computational efficiency of the compressible and the incompressible DG formulations. 

\subsection{Matrix-free implementation of weak forms}
To obtain an efficient compressible Navier--Stokes solver based on high-order DG methods it remains to discuss how to implement equation~\eqref{IntegralsNumericalQuadrature} in an algorithmically and computationally efficient way. In this context, the discretized operators are evaluated by means of matrix-free operator evaluation using sum-factorization, a technique developed by the spectral element community in the 1980s~\cite{Orszag1980}. An important observation is that all terms in equation~\eqref{IntegralsNumericalQuadrature} exhibit the same structure composed of three main steps:
\begin{enumerate}
\item Evaluate the solution (or its gradient in reference coordinates) in all quadrature points by summation over all shape functions (nodes) of an element (or face). This step makes use of the sum-factorization technique in order to reduce operation counts by exploiting the tensor product structure of the shape functions for quadrilateral/hexahedral elements. Here, we examplary show the evaluation of the solution in all quadrature points in 3D
\begin{align}
\forall q: \ \bm{U}^e_h(\bm{\xi}_q) = \sum_{i_1=1}^{k+1} l_{i_1}(\xi_{1,q_1}) \sum_{i_2=1}^{k+1} l_{i_2} (\xi_{2,q_2}) \sum_{i_3=1}^{k+1} l_{i_3} (\xi_{3,q_3}) \bm{U}_{i_1 i_2 i_3}^e \ ,\label{SumFactorization}
\end{align}
where the one-dimensional shape functions~$l_i(\xi)$ are Lagrange polynomials with the LGL nodes as support points. The multi-indices~$(i_1,i_2,i_3)$ and~$(q_1,q_2,q_3)$ are associated to node~$i$ and quadrature point~$q$, respectively, using the tensor product structure of the shape functions and the quadrature rule.
\item Perform operations on all quadrature points~$q$, e.g., apply geometry terms arising from the transformation of integrals from reference space to physical space, evaluate (numerical) fluxes, forcing terms on the right hand side of the equations, or boundary conditions for boundary face integrals.
\item Multiply by the test function (or its gradient in reference coordinates) and perform the summation over all quadrature points. This step has to be done for all test functions of an element (or face). Indeed, this step is very similar to step i), except that the role of quadrature points and interpolation points is interchanged. Accordingly, sum-factorization is used as in step i).
\end{enumerate}
Below, we briefly summarize the main building blocks and distinctive features of our matrix-free implementation and refer to~\cite{Kronbichler2017b} for a detailed discussion.

\paragraph{Generality and design choices} The implementation is generic and offers full flexibility regarding the choice of interpolation points and quadrature points. We explicitly mention that the present implementation of the compressible Navier--Stokes equations does not use a collocation approach, i.e., the interpolation points for the Lagrange polynomials and the quadrature points do not coincide. Accordingly, the mass matrix is block-diagonal  but can be inverted exactly in a matrix-free way~\cite{Kronbichler2016}. We will give evidence in this work that the (inverse) mass matrix operator is memory-bound for our implementation for moderately high polynomial degrees, rendering this approach as efficient as inverting a diagonal mass matrix. Hence, there is no performance advantage in using a collocation approach for the present matrix-free implementation. For a more detailed performance analysis of the mass matrix inversion in comparison to the Runge--Kutta vector operations in the setting of acoustics we refer to~\cite{Schoeder2018}.

\paragraph{Sum-factorization} The sum-factorization kernels are highly optimized. Equation~\eqref{SumFactorization} involves the application of~$d$ kernels, each of which is a dense matrix--matrix product of the 1D shape value matrix with the matrix containing the solution coefficients. When evaluating the gradient of the solution, the number of 1D kernels can be reduced from~$d^2$ to~$2d$ by factoring out the above 1D matrices~$l_i(\xi_q)$ for interpolating the solution onto the quadrature points. This technique can be interpreted as a transformation to a collocation basis and a subsequent evaluation of the gradient in the collocation basis~\cite{Kronbichler2017b}. In addition, the number of arithmetic operations for the 1D kernels can be further reduced by exploiting the symmetry of the nodal shape functions, a technique known as even-odd decomposition~\cite{Kopriva09}. Moreover, the length of the loops over 1D nodes and quadrature points is a template parameter allowing the compiler to produce optimal code, e.g., by loop unrolling.

\paragraph{Vectorization} The computations to be performed in equation~\eqref{IntegralsNumericalQuadrature} are the same for all elements. The only difference is that different elements operate on different degrees of freedom. Hence, the algorithm can be vectorized over several elements according to the SIMD paradigm in order to exploit all levels of parallelism offered by modern computer hardware. The granularity is therefore not one cell but rather a batch of cells that are processed concurrently. Hence, the basic data type is not~\texttt{double} but a small array of doubles, called~\texttt{VectorizedArray<double>} in our implementation. The number of data elements in this array depends on the register width of the CPU, e.g., 4 doubles for AVX2 instruction set extensions and 8 doubles for AVX512. While this strategy increases the pressure on caches due to larger temporary data arrays, it is shown in~\cite{Kronbichler2017b} that this strategy pays off as compared to vectorization within a single element for moderately high polynomial degrees.

\paragraph{Complexity} For cell integrals the number of operations scales as~$\mathcal{O}(k^{d+1})$, i.e., the complexity is linear in~$k$ per degree of freedom. Similarly, face integrals have a complexity of~$\mathcal{O}(k^{d})$ due to the lower dimensionality of faces,~$d_f = d-1$. Solution coefficients have to be stored per degree of freedom and geometry information per quadrature point. Accordingly, the memory requirements and data transfers to/from main memory scale as~$\mathcal{O}(k^d)$, i.e., the complexity is constant per degree of freedom. As a result, it is unclear whether the overall efficiency of the implementation shows a complexity that is constant or linear in~$k$. The behavior depends on the relative costs of cell integrals compared to face integrals as well as the hardware under consideration, i.e., whether the method can be characterized as compute-bound or memory-bound. On modern multi-core CPU architectures, the efficiency of our approach is in fact almost independent of~$k$ for moderately high polynomial degrees~$k\leq 10$, see~\cite{Fehn2018b,Kronbichler2017b}. Numerical evidence for this behavior in the context of the present compressible DG solver is given in Section~\ref{NumericalResults}.

\medskip
The present compressible Navier--Stokes code is implemented in~\texttt{C++} and makes use of the object-oriented finite element library~\texttt{deal.II}~\cite{dealII85} as well as the generic interface for matrix-free evaluation of weak forms~\cite{Kronbichler2012,Kronbichler2017b}. While the focus is on the node-level performance of the matrix-free implementation in the present work, we mention that our approach is also well-suited for massively-parallel computations on modern HPC clusters. For parallel runs on large-scale supercomputers the implementation uses an efficient MPI parallelization where parallel scalability  up to~$\mathcal{O}(10^5)$ cores has been shown in~\cite{Krank2017,Kronbichler2016b}.

\subsection{Temporal discretization}
We consider explicit Runge--Kutta time integration methods for discretization in time~\cite{Kennedy2000,Toulorge2012,Kubatko2014}. The explicit treatment of convective terms and viscous terms sets an upper bound on the time step size in order to maintain stability. As discussed below and thoroughly investigated in this work, the time step limitation related to the viscous term is less restrictive than the convective one for flow problems characterized by low viscosities and coarse spatial resolutions. Moreover, for the type of flow problems considered in this study one typically observes that the time step limitations arising from the CFL condition are restrictive in the sense that significantly larger time steps would be possible from an accuracy point of view and, hence, would be desirable in terms of efficiency. Accordingly, when developing explicit Runge-Kutta time integration methods for such problems, the main aim is to maximize the region of linear stability. In this respect, additional parameters originating from additional stages of the Runge--Kutta method are used to maximize the domain of linear stability. The system of ordinary differential equations arising from discretization in space is
\begin{align}
\frac{\mathrm{d} \underline{U}}{\mathrm{d} t} = \underline{M}^{-1}\left(-\underline{F}_{\mathrm{c}}\left(\underline{U},t\right)-\underline{F}_{\mathrm{v}}\left(\underline{U},t\right)+\underline{F}_{\mathrm{rhs}}\left(\underline{U},t\right)\right) = \underline{M}^{-1}\left(-\underline{F}\left(\underline{U},t\right)+\underline{F}_{\mathrm{rhs}}\left(\underline{U},t\right)\right)
= \underline{L}\left( \underline{U} , t \right) \; .\label{SpatiallyDiscretizedProblem}
\end{align}
Using an explicit,~$s$-stage Runge-Kutta time integration scheme in Butcher notation, the time discrete problem is given as
\begin{align}
\begin{split}
\underline{k}^{(i)} &= \underline{L}\left(\underline{U}_n + \Delta t \sum_{j=1}^{i-1} a_{ij} \underline{k}^{(j)}, t_n + c_i \Delta t  \right) \; , \; \;  i=1,...,s \; ,\\
\underline{U}_{n+1} &= \underline{U}_n + \Delta t \sum_{i=1}^{s} b_i \underline{k}^{(i)}\; ,
\end{split}
\end{align}
with~$c_i = \sum_{j=1}^{s}a_{ij}$ and~$a_{ij} = 0$ for~$j\geq i$. We implemented and tested Runge-Kutta methods developed in~\cite{Kennedy2000,Toulorge2012,Kubatko2014}. While the methods developed in~\cite{Kennedy2000} are generic in the sense that they are not designed for a specific class of discretization methods, the methods proposed in~\cite{Toulorge2012,Kubatko2014} aim at maximizing the critical time step size in view of discontinuous Galerkin discretization methods. Furthermore, the low-storage Runge--Kutta methods developed in~\cite{Toulorge2012} have the advantage that they involve less vector update operations per time step as compared to the strong-stability preserving RK methods (SSPRK) of~\cite{Kubatko2014}. For this reason, and in agreement with our numerical experiments, the low-storage explicit Runge--Kutta methods of~\cite{Toulorge2012} appear to be the most efficient time integration schemes. Hence, we use the third order~$p=3$ low-storage explicit Runge--Kutta method with~$s=7$ stages of~\cite{Toulorge2012} that resulted in the best performance in our preliminary tests, i.e., it allows time step sizes as large as for SSPRK methods but requires less memory and less vector update operations. We strengthen, however, that the differences in performance between these different time integration schemes are rather small and that choosing any other time integration scheme within the class of explicit Runge--Kutta methods mentioned above would not change the conclusions drawn in this work. For the low-storage Runge-Kutta method used in this work, one additional vector has to be stored apart from the solution vectors~$\underline{U}_{n}$ and~$\underline{U}_{n+1}$. 

Both the convective term and the viscous term are integrated explicitly in time, introducing restrictions of the time step size according to~\cite[Chapter 7.5]{Hesthaven07}
\begin{align}
\Delta t_{\mathrm{c}} = \frac{\mathrm{Cr}}{k^{e}}\frac{h_{\mathrm{min}}}{\Vert \bm {u} \Vert_{\mathrm{max}}+c} \; , \; 
\Delta t_{\mathrm{v}} = \frac{\mathrm{D}}{k^{f}}\frac{h_{\mathrm{min}}^2}{\nu} \; ,\label{TimeStepRestrictions_Compressible}
\end{align}
where~$\mathrm{Cr}$ is the Courant number and~$\mathrm{D}$ the respective non-dimensional quantity for the viscous time step restriction. Moreover,~$\Vert \bm {u} \Vert_{\mathrm{max}}$ denotes an estimate of the maximum velocity,~$c=\Vert \bm {u} \Vert_{\mathrm{max}}/\mathrm{M}$ the speed of sound, and~$h_{\mathrm{min}}$ is a characteristic element length scale computed as the minimum vertex distance. In order to guarantee stability, the overall time step size~$\Delta t$ is chosen as the minimum of both time step restrictions,~$\Delta t = \min \left(\Delta t_{\mathrm{c}} , \Delta t_{\mathrm{v}}\right)$. The exponents~$e$ and~$f$ describe the dependency of the time step size as a function of the polynomial degree. Values of~$e=2$ and~$f=4$ are specified in~\cite{Hesthaven07}. However, numerical results shown in Section~\ref{NumericalResults} give evidence that~$e=1.5$ and~$f=3$ model the relation between polynomial degree and maximal possible time step size more accurately for the problem considered here, allowing to use an almost constant~$\mathrm{Cr}$ number or~$\mathrm{D}$ number over a wide range of polynomial degrees~$1\leq k \leq 15$. In our experience,~$e=2$ and~$f=4$ appear to be conservative estimates but we also mention that these values might well depend on the problem or other parameters such as the Reynolds number. From a practical point of view, exact values of~$e$ and~$f$ are desirable to avoid the necessity of re-adjusting the~$\mathrm{Cr}$ number or~$\mathrm{D}$ number for different polynomial degrees, and to perform simulations for different polynomial degrees with maximal computational efficiency. An important question is which of the two time step restrictions is the limiting one
\begin{align}
\frac{\Delta t_{\mathrm{c}}}{\Delta t_{\mathrm{v}}} = \frac{\mathrm{Cr}}{\mathrm{D}} \frac{\Vert \bm {u} \Vert_{\mathrm{max}}}{\Vert \bm {u} \Vert_{\mathrm{max}}+c}\frac{\nu}{\Vert \bm {u} \Vert_{\mathrm{max}} h_{\mathrm{min}}/k^{f-e}} = \frac{\mathrm{Cr}}{\mathrm{D}} \frac{\mathrm{M}}{1+\mathrm{M}}\frac{1}{\mathrm{Re}_{h,k}} \; .
\end{align}
The critical~$\mathrm{Cr}$ number is typically significantly larger than the critical~$\mathrm{D}$ number,~$\mathrm{Cr}_{\mathrm{crit}}/\mathrm{D}_{\mathrm{crit}}\gg 1$ and the Mach number is~$\mathrm{M}=\mathcal{O}(10^{-1})$ for the low-Mach number problems considered in the present work. Accordingly, we expect that the factor in front of the element Reynolds number is approximately~$\mathcal{O}(1)$. For highly under-resolved situations, the element Reynolds number is~$\mathrm{Re}_{h,k} \gg 1$. In this regime, one can expect that the convective time step size is more restrictive than the viscous time step size. Only for well-resolved situations which are numerically viscous dominated,~$\mathrm{Re}_{h,k} \leq 1$, the viscous term is expected to limit the time step size according to the above relation. Since this is typically only the case for computations with DNS-like resolutions, one can expect the convective time step restriction to be the relevant one for strongly under-resolved LES computations. Apart from these theoretical considerations, we will verify these assumptions by means of numerical investigations in Section~\ref{NumericalResults}.

The computational costs per time step mainly originate from the evaluation of the nonlinear compressible Navier--Stokes operator, equation~\eqref{SpatiallyDiscretizedProblem}, that has to be evaluated once within every stage of the Runge-Kutta method. For Runge--Kutta methods with~$s$ stages, we therefore define the quantity
\begin{align}
\Delta t_{s} = \frac{\Delta t}{s}
\end{align}
to measure the efficiency of the time integration scheme. Similarly, we introduce a Courant numbers that is normalized by the number of Runge--Kutta stages~$s$
\begin{align}
\mathrm{Cr}_{s} = \frac{\mathrm{Cr}}{s} \; .
\end{align}
According to the above relations, a Runge--Kutta scheme with a larger number of stages is only more efficient if it allows to use a larger time step per Runge--Kutta stage, i.e., a larger~$\Delta t_s$ or~$\mathrm{Cr}_{s}$.

\section{Reference method: incompressible Navier--Stokes DG solver}\label{IncompressibleSolver}
As a reference method, we consider a high-performance DG solver for the incompressible Navier--Stokes equations consisting of the momentum equation and the continuity equation
\begin{align}
\frac{\partial \bm{u}}{\partial t} + \nabla \cdot \left(\bm{u}\otimes \bm{u}\right) - \nu \Laplace{\bm{u}} + \nabla p_{\mathrm{k}} &= \bm{f}\; , \label{MomentumEquationIncompressible}\\
\nabla \cdot \bm{u} &= 0 \; ,\label{ContinuityEquationIncompressible}
\end{align}
where~$\bm{u}$ is the velocity,~$p_{\mathrm{k}}$ the kinematic pressure, and~$\bm{f}$ the body force vector. The incompressibility constraint,~$\Div{\bm{u}}=0$, requires discretization techniques in both space and time differing from those used to discretize the compressible Navier--Stokes equations. 

\subsection{Temporal discretization} For discretization in time, our reference incompressible DG solver uses the dual splitting scheme~\cite{Karniadakis1991} 
\begin{align}
\frac{\gamma_0\hat{\bm{u}}-\sum_{i=0}^{J-1}\left(\alpha_i\bm{u}^{n-i}\right)}{\Delta t} &= 
- \sum_{i=0}^{J-1}\left(\beta_i \Div{\left(\bm{u}^{n-i}\otimes\bm{u}^{n-i}\right)}\right)
+ \bm{f}\left(t_{n+1}\right)\; ,\label{DualSplitting_ConvectiveStep}\\
-\nabla^2 p_{\mathrm{k}}^{n+1} &= -\frac{\gamma_0 }{\Delta t}\Div{\hat{\bm{u}}} \; ,\label{DualSplitting_PressureStep}\\
\hat{\hat{\bm{u}}} &= \hat{\bm{u}} - \frac{\Delta t}{\gamma_0} \Grad{p_{\mathrm{k}}^{n+1}}\; ,\label{DualSplitting_ProjectionStep}\\
\frac{\gamma_0 }{\Delta t} \bm{u}^{n+1}  -  \nu\Laplace{\bm{u}^{n+1}} &=
\frac{\gamma_0 }{\Delta t}\hat{\hat{\bm{u}}} \; .\label{DualSplitting_ViscousStep}
\end{align}
The dual splitting scheme belongs to the class of projection methods~\cite{Guermond06} which are widely used to obtain efficient time integration algorithms for incompressible flows. Projection methods have in common that a Poisson equation has be solved for the pressure, see equation~\eqref{DualSplitting_PressureStep}, and that a divergence-free velocity is obtained by projection, equation~\eqref{DualSplitting_ProjectionStep}. For this reason, the viscous term is typically formulated implicitly, equation~\eqref{DualSplitting_ViscousStep}, to avoid time step restrictions arising from the discretization of terms involving second derivatives. Note that one has to solve a linear system of equations -- the pressure Poisson equation -- in every time step anyway. Hence, formulating the viscous term implicitly increases the costs per time step only moderately. The convective term is treated explicitly in case of the dual-splitting scheme, equation~\eqref{DualSplitting_ConvectiveStep}, so that this scheme can be denoted as a semi-implicit or mixed implicit-explicit method with respect to the formulation of the viscous term and the convective term. As a consequence of formulating the convective term explicitly, the time step size is restricted according to the CFL condition. Adopting the notation in~\cite{Fehn2018b} we obtain
\begin{align}
\Delta t_{\mathrm{inc}} = \frac{\mathrm{Cr}_{\mathrm{inc}}}{k^{e}}\frac{h_{\mathrm{min}}}{\Vert \bm {u} \Vert_{\mathrm{max}}} \; ,\label{CFL_Condition}
\end{align}
where~$e=1.5$ was found to be a tight fit for the incompressible Navier--Stokes DG solver~\cite{Fehn2018b}. The Mach number does not occur in the incompressible case, leading to significantly larger time step sizes for the incompressible solver as discussed in Section~\ref{EfficiencyModels}.

\subsection{Spatial discretization} The discretization of the above equations in space is based on high-order discontinuous Galerkin methods similar to the compressible Navier--Stokes solver. The Lax--Friedrichs flux is used for the convective term and the symmetric interior penalty method for the viscous term and the pressure Poisson operator. Central fluxes are used for the discretization of the velocity divergence term and the pressure gradient term, see also~\cite{Fehn2017} for a detailed discussion regarding the DG discretization of these terms in the context of splitting methods. Additionally, consistent penalty terms enforcing the continuity equation are applied in the projection step, which were found to be essential components to render the DG discretization method a robust discretization scheme in the regime of under-resolved turbulent flows~\cite{Fehn18}. Our DG discretization makes use of fully-discontinuous function spaces for velocity and pressure unknowns using polynomial shape functions of one degree higher for the velocity than for the pressure for reasons of inf--sup stability. From the point of view of finite element function spaces, the use of penalty terms might also be interpreted as a realization of divergence-free~$H(\mathrm{div})$ elements in a weak sense. For reasons of brevity, we do not specify the details of the spatial discretization approach but refer to~\cite{Fehn2017,Fehn18} and references therein for a detailed discussion of these aspects. Standard quadrature with~$k_{u|p}+1$ quadrature points is used for numerical integration except for the convective term for which we use a~$3/2$-dealiasing strategy due to the quadratic nonlinearities of the convective term.

\subsection{Fully-discrete problem} After discretization in space, the following sub-steps have to be solved within each time step to obtain the solution~$\underline{u}^{n+1}$ and~$\underline{p}_{\mathrm{k}}^{n+1}$~\cite{Fehn18}
\begin{align}
\frac{\gamma_0 \hat{\underline{u}}-\sum_{i=0}^{J-1}\left(\alpha_i\underline{u}^{n-i}\right)}{\Delta t}
&= \underline{M}^{-1}\left(
- \sum_{i=0}^{J-1} \left(\beta_i \underline{C}\left(\underline{u}^{n-i}\right) \right)
+ \underline{f}(t_{n+1})\right) \; ,
\label{DualSplitting_ConvectiveStep_MatrixForm}\\
\underline{L}_{\text{hom}}\;\underline{p}_{\mathrm{k}}^{n+1} &= - \frac{\gamma_0}{\Delta t} \underline{D}\; \hat{\underline{u}} - \underline{L}_{\text{inhom}}
\; ,
\label{DualSplitting_Pressure_MatrixForm}\\
\left(\underline{M} + \underline{A}_{\mathrm{D}} + \underline{A}_{\mathrm{C}} \right)\hat{\hat{\underline{u}}}  &= \underline{M}\;\hat{\underline{u}}-\frac{\Delta t}{\gamma_0}\underline{G}\;\underline{p}_{\mathrm{k}}^{n+1}\; ,\label{DualSplitting_Projection_MatrixForm}\\
\left(\frac{\gamma_0}{\Delta t} \underline{M}
+ \underline{V}_{\mathrm{hom}}\right)\underline{u}^{n+1}
&= 
\frac{\gamma_0}{\Delta t}\underline{M}\;\hat{\hat{\underline{u}}}-\underline{V}_{\mathrm{inhom}}
\; ,
\label{DualSplitting_ViscousStep_MatrixForm}
\end{align}
While the first sub-step involves operations similar to the explicit compressible Navier--Stokes solver presented above, a linear system of equations has to be solved in each of the remaining three sub-steps. Efficient preconditioning strategies for these symmetric, positive-definite problems in the context of high-order DG discretizations have been developed in~\cite{Krank2017} that are completely based on matrix-free operations and that achieve optimal complexity with respect to mesh refinement. While our approach uses the inverse mass matrix as preconditioner for the projection step and the visous step, we apply geometric multigrid methods with efficient matrix-free smoothing to obtain mesh-independent convergence also for the pressure Poisson equation. It is worth noting that the inverse mass matrix operation is realized in a matrix-free way~\cite{Kronbichler2016}. In terms of compute performance, this operation is a memory-bound operation so that applying the inverse mass matrix is as expensive as applying a diagonal mass matrix although the mass matrix is block-diagonal and is inverted exactly. Let us emphasize that realizing all solver components in a matrix-free way is an essential prerequisite to allow a fair comparison with the explicit compressible DG solver.

\section{Discussion of efficiency of compressible and incompressible flow solvers}\label{EfficiencyModels}
\subsection{Efficiency model} 
Following~\cite{Fehn2018b}, we define the efficiency~$E$ of a numerical method as
\begin{align}
E = \frac{\text{accuracy}}{\text{computational costs}} = \frac{\text{accuracy}}{\text{wall time}\cdot \text{number of cores}} \; ,
\end{align}
i.e., a method is more efficient if it obtains a more accurate solution within a given amount of computational costs or if it allows to reduce computational costs reaching the same level of accuracy. Moreover, the product of wall time and the number of cores used for a simulation is defined as a measure of computational costs. From this macroscopic point of view it is difficult to trace differences in efficiency between the compressible and the incompressible flow solver. Hence, we factor out the above equation into the main algorithmic components of the numerical method for the compressible flow solver on the one hand and the incompressible flow solver on the other hand.

For the compressible DG solver with explicit Runge--Kutta time integration, each time step consists of several Runge--Kutta stages where the computational costs per Runge--Kutta stage mainly stem from the evaluation of the spatially discretized compressible Navier--Stokes operator
\begin{align}
\begin{split}
E_{\mathrm{comp}} = \underbrace{\frac{\text{accuracy}}{\text{DoFs}}}_{\text{spatial discretization}}
\cdot \underbrace{\frac{1}{\text{time steps}\cdot \text{RK stages}}}_{\text{temporal discretization}} \cdot
\underbrace{\frac{\text{DoFs}\cdot\text{time steps}\cdot\text{RK stages}}{\text{computational costs}}}_{\text{implementation}}  \; .
\end{split}
\end{align}
The above equation identifies the spatial discretization, the temporal discetization, and the implementation as the factors that determine the overall efficiency of the numerical method. The spatial discretization approach is efficient if it is accurate for a given number of unknowns. Similarly, an efficient time integration method is characterized by a large stability region allowing to use a large time step~$\Delta t_s$ per Runge--Kutta stage. The last factor quantifies how many degrees of freedom can be processed/updated per time step and per Runge--Kutta stage within a given amount of computational costs. We denote this quantity also as throughput of the matrix-free evaluation of discretized operators. In the literature, the performance index (PID) is used to quantify the efficiency of the implementation, see~\cite{Beck2018}. The performance index PID~\footnote{We mainly use the efficiency of the implementation in terms of degrees of freedom processed per second per core instead of the performance index PID, since the performance index would be larger for less efficient implementations and since the throughput directly occurs as a multiplicative factor in the overall efficiency of the method. Accordingly, measuring the efficiency in terms of~$\mathrm{DoFs}/(\mathrm{sec}\cdot\mathrm{core})$ seems to be a more intuitive definition.} is defined as the time needed to update the $d+2$ unknowns corresponding to one scalar degree of freedom on one core per time step and per Runge--Kutta stage
\begin{align*}
\mathrm{PID} = \frac{\text{wall time}\cdot \text{number of cores}}{\text{scalar DoFs}\cdot\text{time steps}\cdot\text{RK stages}} = \frac{\text{computational costs}}{\text{scalar DoFs}\cdot\text{time steps}\cdot\text{RK stages}} \; .
\end{align*}
Note that the number of scalar degrees of freedom refers to the number of nodes of the DG discretization in the above definition, i.e., it does not include the factor~$d+2$ for the number of conserved variables.

In case of the incompresssible Navier--Stokes solver and in contrast to the compressible flow solver, each time step is composed of several substeps than can be solved explicitly (convective step) or for which a linear system of equations has to be solved (pressure, projection, and viscous steps). Since the convective term has to be evaluated only once within each time step, this step is negligible in terms of costs~\cite{Fehn2018b} and the efficiency of the incompressible solver depends on how fast one can solve the linear systems of equations in the remaining three sub-steps
\begin{align}
\begin{split}
E_{\mathrm{inc}} = \underbrace{\frac{\text{accuracy}}{\text{DoFs}}}_{\text{spatial discretization}}
\cdot \underbrace{\frac{1}{\text{time steps}}}_{\text{temporal discretization}}
\cdot
\underbrace{\frac{1}{\text{iterations}}}_{\text{iterative solvers}}
\cdot
\underbrace{\frac{\text{DoFs}\cdot\text{time steps}\cdot\text{iterations}}{\text{computational costs}}}_{\text{implementation}} \; .
\end{split}
\end{align}
Due to several operators involved in case of the incompressible flow solver, see equations~\eqref{DualSplitting_ConvectiveStep_MatrixForm},~\eqref{DualSplitting_Pressure_MatrixForm},~\eqref{DualSplitting_Projection_MatrixForm},~\eqref{DualSplitting_ViscousStep_MatrixForm}, the efficiency of the implementation (throughput) can not be expressed as a single number as for the compressible solver. The factor labeled iterations does not only include evaluations of the discretized operators but also preconditioners of different complexity (inverse mass matrix versus geometric multigrid) so that this quantity should be understood as an effective number of operator evaluations per time step, see also the discussion in~\cite{Fehn2018b}. We can now discuss expected differences in efficiency between the compressible and the incompressible flow solvers serving as a model for the efficiency measurements performed in Section~\ref{NumericalResults}.

\subsection{Spatial discretization} 
With respect to the efficiency of the spatial discretization we expect the differences between both solvers to be small. High-order discontinuous Galerkin discretizations with Lax--Friedrichs flux for the convective term and the symmetric interior penalty method for viscous terms are expected to show a similar performance for both the compressible and the incompressible solver. However, to draw precise conclusions a numerical investigation of this aspect has to be performed.

\subsection{Temporal discretization} 
The temporal discretization or, more precisely, time step restrictions arising from an explicit treatment of the convective term (and the viscous term) pose a major difference between the compressible solver and the incompressible solver. Here, we assume that the convective term results in a more restrictive time step size than the viscous term in case of the compressible solver. This assumption is valid for the flow problems considered in this work and will be verified in Section~\ref{NumericalResults}. The time integration methods used for both solvers have different stability regions resulting in different critical~$\mathrm{Cr}$ numbers. More importantly, the CFL condition involves the speed of sound (or the Mach number) in case of the compressible solver
\begin{align}
\frac{\Delta t_{s,\mathrm{comp}}}{\Delta t_{\mathrm{inc}}} = \frac{\frac{\mathrm{Cr}_{\mathrm{comp}}}{s\; k^e} \frac{h_{\mathrm{min}}}{\Vert \bm {u} \Vert_{\mathrm{max}}+c}}{\frac{\mathrm{Cr}_{\mathrm{inc}}}{k^e}\frac{h_{\mathrm{min}}}{\Vert \bm {u} \Vert_{\mathrm{max}}}}=
\frac{\mathrm{Cr}_{s,\mathrm{comp}}}{\mathrm{Cr}_{\mathrm{inc}}}  \frac{\Vert \bm {u} \Vert_{\mathrm{max}}}{\Vert \bm {u} \Vert_{\mathrm{max}}+c}
=\frac{\mathrm{Cr}_{s,\mathrm{comp}}}{\mathrm{Cr}_{\mathrm{inc}}} \frac{\mathrm{M}}{1+\mathrm{M}} 
\overset{\mathrm{M}\ll 1}{\approx} \frac{\mathrm{Cr}_{s,\mathrm{comp}}}{\mathrm{Cr}_{\mathrm{inc}}} \mathrm{M} \; .\label{RatioTimeStepSizesCompInc}
\end{align}
For small Mach numbers ($\mathrm{M}\approx 0.1$ is used in this work to mimic incompressible flows) this results in a significantly smaller time step size for the compressible solver assuming that~$\mathrm{Cr}_{s,\mathrm{comp}}\leq \mathrm{Cr}_{\mathrm{inc}}$. The term on the right-hand side of the above equation poses a potential performance penalty of the compressible solver. In Section~\ref{NumericalResults}, we determine the critical Courant numbers numerically for both solvers and the respective time integration methods.

\subsection{Iterative solution of systems of equations} 
Apart from the time step limitations discussed above, the need to solve linear systems of equations when using the incompressible solver is the second major algorithmic difference to the compressible solver. We explicitly mention, however, that all solver components and preconditioners in case of the incompressible solver are realized in a matrix-free way using the same implementation kernels to allow a fair comparison to the compressible DG solver. Moreover, we use an explicit treatment of the convective term for both solution approaches. Of course, fully implicit formulations constitute a potential performance improvement but this topic is beyond the scope of the present study.

\subsection{Throughput of matrix-free operator evaluation} 
The incompressible solver involves comparably simple operators such as the scalar Laplace operator in case of the pressure Poisson equation. The compressible solver involves significantly more complex operators which can be seen by comparing equations~\eqref{CompressibleNavierStokesCompact} and~\eqref{CompressibleNavierStokesFluxes} for the compressible solver to equations~\eqref{DualSplitting_ConvectiveStep},~\eqref{DualSplitting_PressureStep},~\eqref{DualSplitting_ProjectionStep},~\eqref{DualSplitting_ViscousStep} for the incompressible solver. For this reason, we expect the compressible flow solver to reach a lower throughput for the matrix-free evaluation of finite element operators as compared to the incompressible solver. Again, it is difficult to quantify this difference from a theoretical point of view. Instead, we prefer a detailed numerical investigation of this aspect in Section~\ref{NumericalResults}. As will be shown in the following, differences in throughput have a significant impact on the relative performance of both solvers.

To avoid aliasing effects~\cite{Hesthaven07}, we use an over-integration strategy, i.e., an increased number of quadrature points for the numerical integration of nonlinear terms in the weak formulation, in order to obtain nonlinear stability. For the incompressible flow solver with an explicit treatment of the convective term, this potential decrease in compute performance is practically not visible since the computational costs of the convective step are small as compared to the other sub-steps of the splitting scheme. On the contrary, the increased complexity due to an increased number of quadrature points directly impacts the performance of the compressible DG solver. Theoretically, the complexity increases by a factor of~$\left(n_{q,\mathrm{over}}/({k+1})\right)^d$ in case of over-integration with~$n_{q,\mathrm{over}}$ quadrature points compared to standard integration with~$k+1$ points. The impact of overintegration on the throughput of the matrix-free implementation is also subject of our numerical investigations in Section~\ref{NumericalResults}.

\section{Numerical results}\label{NumericalResults}
As mentioned above, many aspects discussed in this work can only be analyzed by means of numerical investigation. We extensively document performance results for the proposed compressible high-order DG solver and thoroughly compare the efficiency of this approach in relation to an incompressible high-order DG solver based on the same implementation framework. The considered test cases are well-established benchmark problems for transitional and turbulent flows, the 3D Taylor--Green vortex problem and turbulent channel flow.

\subsection{Efficiency of matrix-free operator evaluation}\label{ResultsMatrixFreeImplementation}
In this section, we quantify the efficiency of the matrix-free implementation for the compressible Navier--Stokes equations in terms of degrees of freedom processed per second for several operators. We investigate the convective operator~$\underline{F}_{\mathrm{c}}$ in equation~\eqref{SpatiallyDiscretizedProblem}, the viscous operator~$\underline{F}_{\mathrm{v}}$ in equation~\eqref{SpatiallyDiscretizedProblem}, and the combination of both the convective operator and the viscous operator~$\underline{F}$ in equation~\eqref{SpatiallyDiscretizedProblem}. Apart from the convective and viscous operators, the inverse mass matrix operator~$\underline{M}^{-1}$ and vector update operations such as~$\underline{U} = \alpha \underline{U} + \beta \underline{V}$ are important operations in the explicit Runge--Kutta time integrator that are considered in the following. Finally, the whole operator~$\underline{L}$ on the right-hand side of equation~\eqref{SpatiallyDiscretizedProblem} is investigated. 

All performance measurements are performed on SuperMUC Phase 2 in Garching, Germany. This system is based on Intel Haswell processors (Intel Xeon CPU E5-2697 v3) with the AVX2 instruction set extension. A node consists of 28 cores running at~$2.6$ GHz and the bandwidth to main memory per node is~$137$ GByte/sec. The simulations are run on one fully loaded node by using~$28$ MPI ranks. As usual, a Cartesian mesh with periodic boundary conditions is considered to measure the performance of the matrix-free implementation unless specified otherwise. Additionally, we also discuss the performance on general unstructured meshes.
To make sure that the solution vectors have to be streamed from main memory and can not reside in the cache, the refinement level of the uniform Cartesian grid is~$l=7$ for~$k=1$,~$l=6$ for~$k=2,3$,~$l=5$ for~$k=4,...,7$, and~$l=4$ for~$k=8,...,15$, resulting in~$(1.5 - 8.4)\cdot 10^7$ degrees of freedom. For example, approximately~$0.9 \cdot 10^7$ double precision values would fit into the L3 cache of size~$70$~MB. The measured wall time~$t_{\mathrm{wall}}$ is the wall time averaged over 100 operator evaluations and the minimum wall time out of ten consecutive runs is used.

\subsubsection{Standard quadrature with~$k+1$ quadrature points}
\begin{figure}
 \centering 
\includegraphics[width=1.0\textwidth]{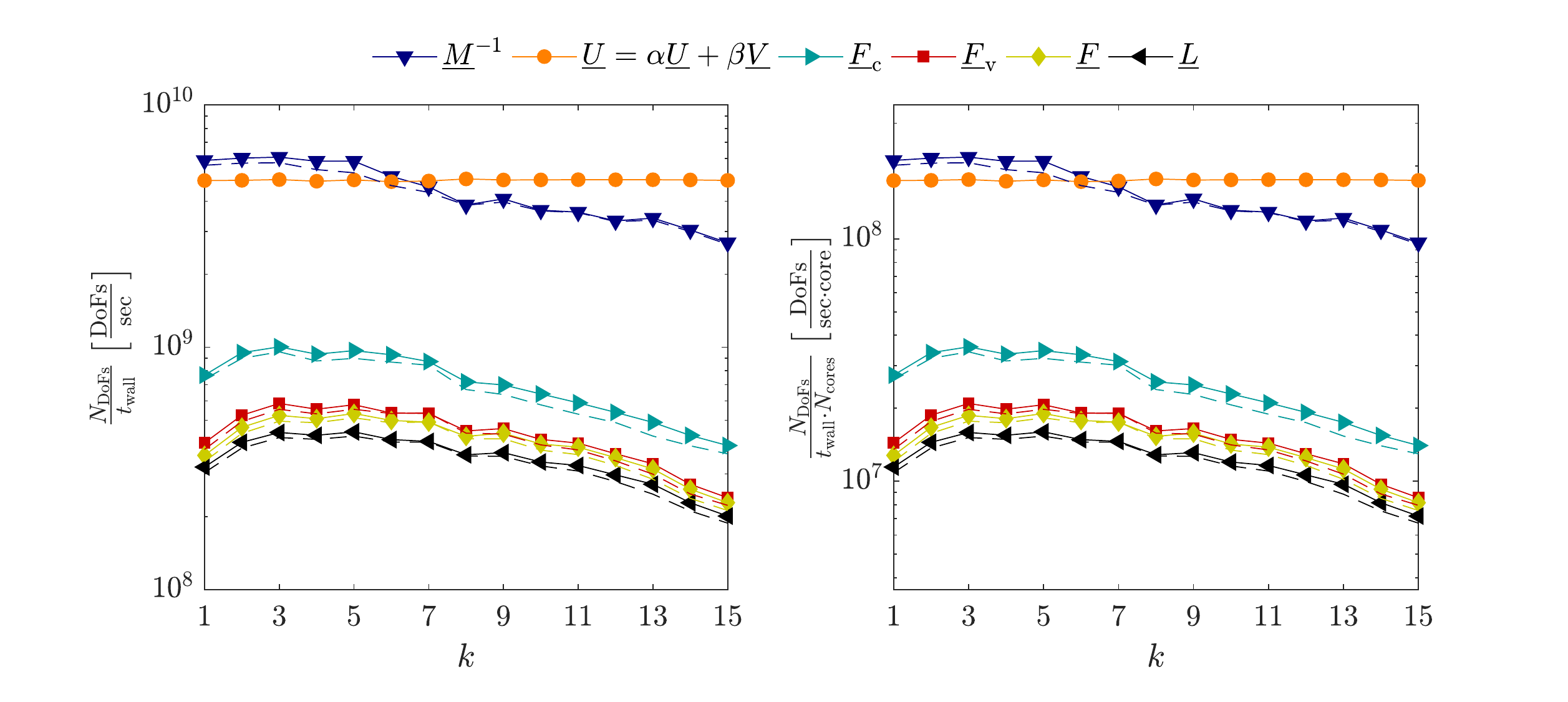}
\caption{Efficiency of matrix-free operator evaluation on a Cartesian mesh in 3D with~$k+1$ quadrature points (standard quadrature). Performance results obtained on a non-Cartesian mesh are shown as dashed lines for comparison. The performance measurements are performed on a fully-loaded node with 28 cores. The throughput in DoFs/sec is shown on the left and the efficiency of the implementation in DoFs/(sec~$\cdot$ core) on the right.}
\label{fig:performance_matrix_free_cartesian_std_quad}
\end{figure}
\begin{figure}[!ht]
 \centering 
 \subfigure[Over-integration with~$n_{\mathrm{q}}=\lceil \frac{3k+1}{2}\rceil$ quadrature points compared to standard quadrature (dashed lines) with~$k+1$ quadrature points.]
 {\includegraphics[width=1.0\textwidth]{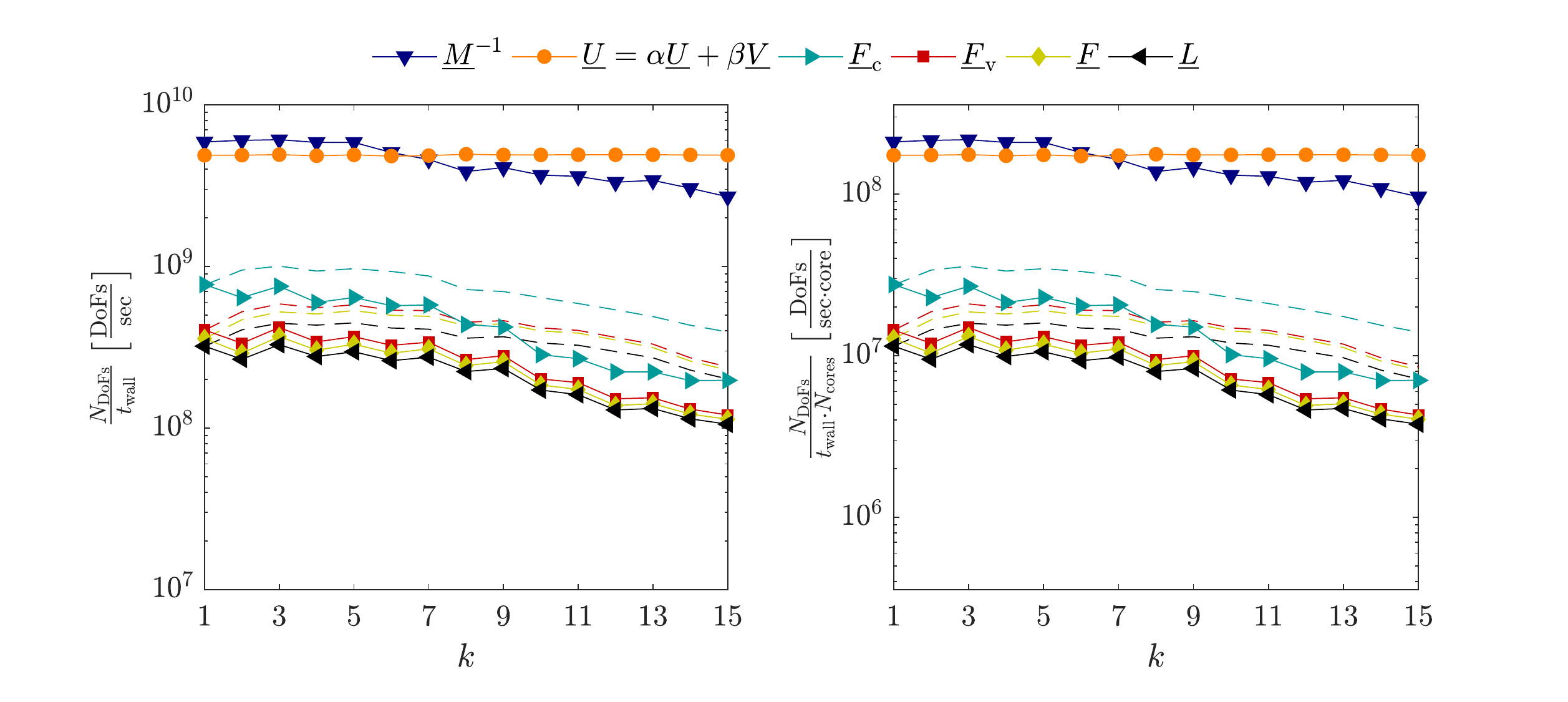}}
 \subfigure[Over-integration with~$n_{\mathrm{q}}=\lceil \frac{4k+1}{2}\rceil=2k+1$ quadrature points compared to standard quadrature (dashed lines) with~$k+1$ quadrature points.]
 {\includegraphics[width=1.0\textwidth]{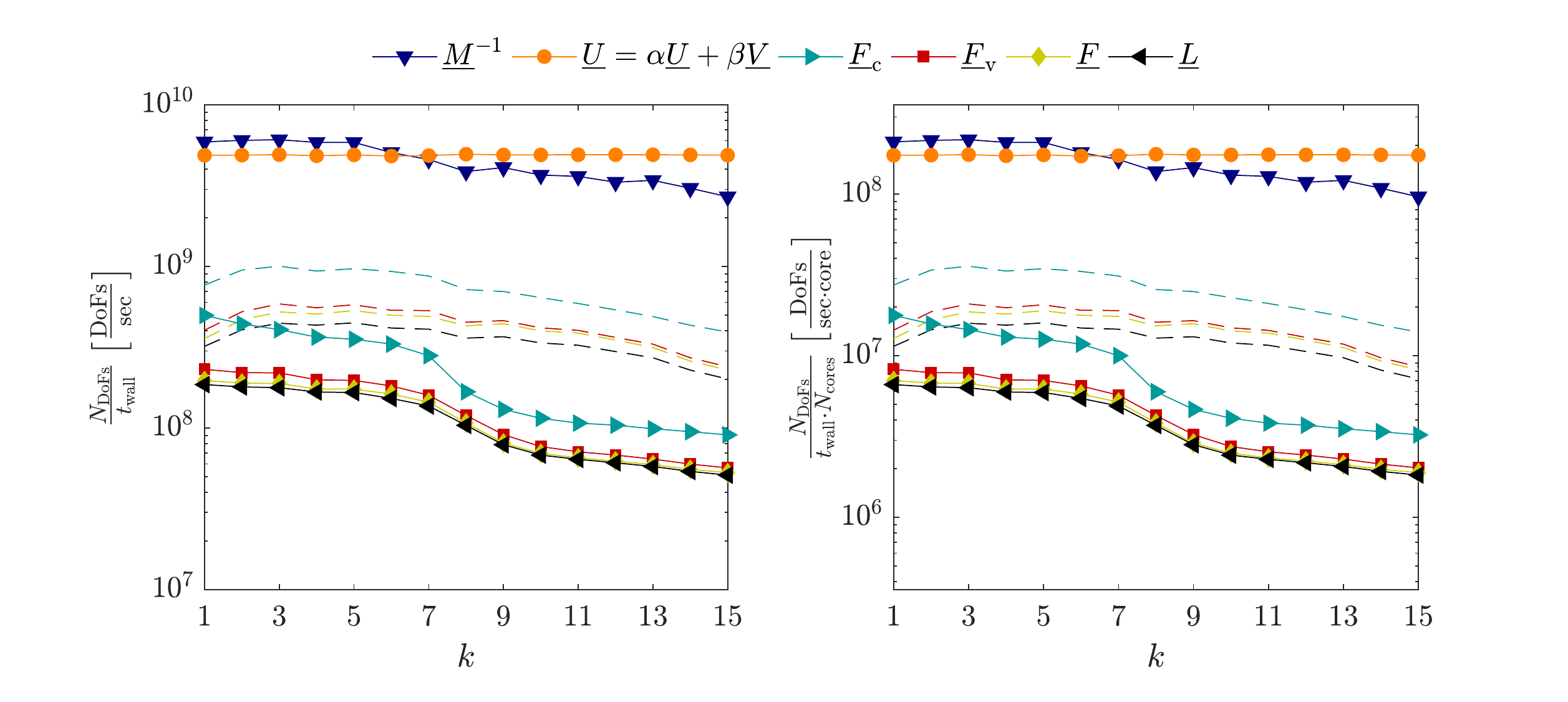}}
\caption{Efficiency of matrix-free operator evaluation on a Cartesian mesh in 3D: Over-integration vs.~standard quadrature (the performance of the inverse mass matrix operator with~$k+1$ quadrature points as well as the vector update operation are shown for comparison).}
\label{fig:efficiency_standard_versus_overintegration}
\end{figure}
In a first step, we investigate the efficiency of the matrix-free implementation for a standard quadrature rule with~$n_{\mathrm{q}}=k+1$ quadrature points in order to allow a direct comparison to the performance of the matrix-free kernels of the incompressible Navier--Stokes DG solver published in~\cite{Fehn2018b}. In Figure~\ref{fig:performance_matrix_free_cartesian_std_quad}, performance measurements are shown for the six operations mentioned above in terms of throughput in DoFs/sec and in terms of the efficiency of the implementation in DoFs/(sec~$\cdot$ core). The vector update operation is memory-bound (only one floating point operation per double precision value transferred to/from main memory), achieving a throughput of approximately~$4.9\cdot 10^9$ DoFs/sec which corresponds to~$85\%$ of the theoretical memory bandwidth of~$137$ GByte/sec. The inverse mass matrix operator is also memory-bound for moderate polynomial degrees~$k\leq 5$. The input vector and the output vector are the same in case of the inverse mass matrix operator,~$\underline{U}=\underline{M}^{-1}\underline{U}$, so that two vectors have to be read from or written to memory achieving a throughput of approximately~$6.0\cdot 10^9$ DoFs/sec which is even higher than for the vector update operation. The throughput reaches approximately~$70\%$ of the theoretical memory bandwidth for low and moderate polynomial degrees. For very high polynomial degrees the performance decreases slightly due to the theoretical complexity of cell integrals of order~$\mathcal{O}(k^{d+1})$ so that the operator tends to become more and more compute bound. The convective operator achieves a throughput of up to~$1 \cdot 10^9$ DoFs/sec for polynomial degrees in the range~$k=3,...,6$. For the viscous operator, the maximum throughput is close to~$6 \cdot 10^8$~DoFs/sec. The reduced throughput compared to the convective term can be explained by the fact that the viscous term involves more expensive terms in the weak form than the convective term, e.g., both the values and the gradients of the solution have to be evaluated for the viscous term, while only the values of the solution are needed for the convective term. The throughput for the combined operator~$\underline{F}$ is very close to the performance of the viscous operator. Scaling the vector by the factor~$-1$ and applying the inverse mass matrix operator to evaluate~$\underline{L}$ introduces almost no additional costs, so that the overall operator~$\underline{L}$ reaches a similar performance with a throughput of up to~$4.5\cdot 10^8$~DoFs/sec.

Compared to the performance results for the matrix-free kernels of the incompressible Navier--Stokes splitting solver published in~\cite{Fehn2018b}, the throughput is in a similar range for the present compressible Navier--Stokes DG solver when using standard quadrature with~$n_{\mathrm{q}}=k+1$ quadrature points. The throughput is somewhat higher for the incompressible Navier--Stokes solver which can be explained by the fact that the dual splitting scheme used for the incompressible solver consists of comparably simple operators while the compressible Navier--Stokes operator involves more complex operations applied to a system of~$d+2$ coupled equations, as already mentioned above.

Evaluating the operator on more complex meshes with deformed elements increases the memory transfer compared to the Cartesian case since the Jacobian matrix is stored for all quadrature point and has to be read from memory. However, since the compressible Navier--Stokes operator is mainly compute bound, evaluating the operator on an unstructured mesh has almost no impact on the performance. This is illustrated in Figure~\ref{fig:performance_matrix_free_cartesian_std_quad} where the efficiency of the matrix-free evaluation for deformed element geometries is shown as dashed lines in addition to the results on the Cartesian mesh. The performance is almost the same for the Cartesian mesh and the unstructured mesh. This might be seen as a positive side effect of the more complex compressible DG operators compared to the incompressible case. For example, the scalar Laplace operator applied during the solution of the pressure Poisson equation in case of the incompressible Navier--Stokes solver already tends to be memory-bound on complex meshes for moderate polynomial degrees for the present matrix-free implementation~\cite{Kronbichler2017b}. Since it can be expected that the peak performance will increase relative to the memory bandwidth on future hardware~\cite{Hager2010}, e.g., AVX512 enabling 8-wide vectorization for double precision values, this effect might improve the performance of the compressible solver relative to the incompressible solver on future hardware.

\subsubsection{Impact of over-integration on performance}\label{PerformanceOverintegration}

Next, we analyze the performance of the matrix-free implementation for an increased number of quadrature points (over-integration). We consider a~$3/2$-dealiasing strategy with~$n_{q}=\lceil \frac{3k+1}{2}\rceil$ quadrature points for quadratic nonlinearities and a more expensive quadrature rule with~$n_{q}=\lceil \frac{4k+1}{2}\rceil=2k+1$ quadrature points for cubic nonlinearities. Results are shown in Figure~\ref{fig:efficiency_standard_versus_overintegration}. The saw tooth behavior visible for the~$3/2$-dealiasing strategy is due to the ratio of quadrature points to the number of degrees of freedom when taking the integer part of~$3k/2$. For~$3/2$-dealiasing, the efficiency is reduced to~$60-70$~\% of the efficiency with standard quadrature for degrees~$2 \leq k\leq 9$, and to~$40-50$~\% for~$k\geq 10$. Using over-integration with~$n_{q}=2k+1$ quadrature points reduces the efficiency to~$30-40$~\% of the efficiency with standard quadrature for degrees~$2 \leq k\leq 7$, and to~$20-25$~\% for~$k\geq 8$. The drop in efficiency particularly visible for~$n_{q}=2k+1$ quadrature points at polynomial degrees~$k=8$ (and also identifiable for 3/2-dealiasing at~$k=10$ for the convective term) can be explained by the fact that the temporary data arrays required for the cell integrals do no longer fit into the L3 cache. The memory required to store the data arrays for the cell integrals can be estimated as~$n_{q}^d\cdot (d+2)\cdot (d+1)\cdot 4 \cdot 8$, where the factors correspond to the number of quadrature points~$n_{q}^d$,~$d+2$ solution fields for the compressible Navier--Stokes equations to be stored for each quadrature point, the fact that both the values and the gradients (factor~$d+1$) are needed to evaluate the convective and viscous terms, and the fact that our code is vectorized over elements with~$4$ double precision values (of~$8$ byte each) fitting into the AVX2 registers. Accordingly, the critical value of~$2.5$~MByte of cache size per processor is reached at polynomial degree~$k=15$ in case of standard integration, at~$k=10$ for 3/2-dealiasing, and at~$k=8$ for~$2k+1$ quadrature points. As a result, some of the temporary data is evicted to main memory and has to be read again causing a significant increase in data transfer to/from main memory and a decrease in throughput. Although over-integration reduces the efficiency, we note that the slowdown due to over-integration is less severe than the theoretical value~$\left(n_{q,\mathrm{over}}/({k+1})\right)^d$. This might be explained by the fact that the sum-factorization steps play a decisive role for which the complexity does not increase with~$\left(n_{q,\mathrm{over}}/({k+1})\right)^d$. We expect that the slowdown due to over-integration is even less severe on future hardware characterized by an increased Flop/Byte ratio~\cite{Hager2010}. Efforts have been made in the past to avoid costly over-integration, e.g., by using filtering~\cite{Gassner2013,Flad2016}. These measures can be expected to cause a loss of accuracy as compared to consistent over-integration, see for example~\cite{Gassner2013}, which has to be taken into account when evaluating the overall efficiency of a discretization scheme. From a pratical point of view, a~$3/2$-dealiasing strategy is often sufficient, see for example~\cite{Beck2014b} and Section~\ref{TaylorGreenVortex} below, and causes an increase in costs by at most a factor of roughly~$1.5$ for moderately high polynomial degrees according to our results. Hence, using over-integration along with an efficient implementation appears to be highly competitive. As for the standard quadrature rule with~$n_{q}=k+1$ quadrature points, the throughput reduces only slightly on complex meshes with deformed elements.

\begin{table}[!h]
\caption{Performance index (PID) in~$10^{-6} \frac{\mathrm{sec}\cdot \mathrm{core}}{\mathrm{DoF}}$ for present matrix-free implementation of the compressible Navier--Stokes equations for different quadrature rules and for selected polynomial degrees~$k=2,3,5,7,11,15$.}\label{tab:Performance_Index}
\renewcommand{\arraystretch}{1.1}
\begin{center}
\begin{tabular}{ccccccc}
\hline
& $k=2$ & $k=3$ & $k=5$ & $k=7$ & $k=11$ & $k=15$\\
\hline
$n_q=k+1$ 						  & 0.35 & 0.31 & 0.31 & 0.34 & 0.43 & 0.70\\
$n_q=\lceil \frac{3k+1}{2}\rceil$ & 0.53 & 0.43 & 0.47 & 0.51 & 0.87 & 1.33\\
$n_q=2k+1$ 						  & 0.78 & 0.79 & 0.84 & 1.02 & 2.19 & 2.74\\
\hline
\end{tabular}
\end{center}
\renewcommand{\arraystretch}{1}
\end{table}

Finally, we compare absolute performance numbers for our implementation to results published in recent literature. In~\cite{Beck2018}, a performance index of~$\mathrm{PID} \approx 1 \cdot 10^{-6} \frac{\mathrm{sec}\cdot \mathrm{core}}{\mathrm{DoF}}$ is achieved for a polynomial degree of~$k=6$ using a collocation approach with~$k+1$ quadrature points. Moreover, a performance index of~$\mathrm{PID}=(1.4-1.5)\cdot 10^{-6} \frac{\mathrm{sec}\cdot \mathrm{core}}{\mathrm{DoF}}$ is specified for a 3/2-dealiasing strategy and a polynomial degree of~$k=7$. The hardware used in~\cite{Beck2018} is also an Intel Haswell system (Intel Xeon CPU E5-2680 v3) as in the present study so the performance numbers should be directly comparable. In Table~\ref{tab:Performance_Index} we list the performance index of the present implementation for a standard quadrature rule with~$k+1$ points and for the over-integration strategies discussed above. Hence, our compressible Navier--Stokes implementation is approximately a factor of~$3$ faster compared to the results published in~\cite{Beck2018}. Once more, these results highlight the importance to use the same implementation for a comparative study to be meaningful. The considerable increase in the performance index (decrease in efficiency) for~$n_q=2k+1$ quadrature points and~$k=11,15$ is again related to the L3 cache behavior discussed above.

%The approach in~\cite{Beck2018} reaches an efficiency of the implementation of~$1\cdot 10^{6} \frac{\mathrm{DoFs}}{\mathrm{sec}\cdot \mathrm{core}}$ for standard quadrature and~$(0.67-0.71) \cdot 10^{6} \frac{\mathrm{DoFs}}{\mathrm{sec}\cdot \mathrm{core}}$ for~3/2-dealiasing.

\subsection{3D Taylor--Green benchmark problem}\label{TaylorGreenVortex}
We analyze the 3D Taylor--Green vortex problem~\cite{Taylor1937} which is a well-established benchmark problem, see for example~\cite{Wang2013}, to assess the accuracy as well as the computational efficiency of turbulent flow solvers.

\subsubsection{Problem description}
The problem is solved on a unit cube~$\Omega_h = [-\pi L,\pi L]^3$ with periodic boundary conditions and is characterized by the following initial condition
\begin{align*}
u_1(\bm{x},t=0) &= U_0\sin\left(\frac{x_1}{L} \right)\cos\left(\frac{x_2}{L}\right)\cos\left(\frac{x_3}{L}\right)\; ,\\
u_2(\bm{x},t=0) &= -U_0\cos\left(\frac{x_1}{L} \right)\sin\left(\frac{x_2}{L}\right)\cos\left(\frac{x_3}{L}\right)\; ,\\
u_3(\bm{x},t=0) &= 0\; ,\\
p(\bm{x},t=0) &= p_0 + \frac{\rho_0 U_0^2}{16}\left(\cos\left(\frac{2x_1}{L}\right)+ \cos\left(\frac{2x_2}{L}\right)\right)\left(\cos\left(\frac{2x_3}{L}\right)+2 \right)\; .
\end{align*}
The parameters are chosen as in~\cite{Wang2013}: The Reynolds number is~$\mathrm{Re}={\rho_0 U_0 L}/{\mu}=1600$, the Prandtl number is~$\mathrm{Pr}={\mu c_p}/{\lambda}=0.71$, the ratio of specific heats is~$\gamma={c_p}/{c_v}=1.4$, and the Mach number is~$\mathrm{M}={U_0}/{c_0}=0.1$. Setting~$\rho_0=1$,~$U_0=1$,~$L=1$, and~$\mathrm{R}=287$, the dynamic viscosity~$\mu$ and thermal conductivity~$\lambda$ can be obtained from the above equations. A uniform temperature~$T(\bm{x},t=0)=T_0$ with~$T_0 = {c^2_0}/{(\gamma \mathrm{R})}$ is presribed, so that the pressure constant is given as~$p_0 = \rho_0 \mathrm{R} T_0$. The density is then given as~$\rho(\bm{x},t=0) = {p(\bm{x},t=0)}/{(\mathrm{R}T_0)}$, and the energy as~$E(\bm{x},t=0) = c_{v} T(\bm{x},t=0) + \bm{u}^{2}(\bm{x},t=0) / 2$. The problem is simulated over the time interval~$0\leq t \leq t_{\mathrm{f}}=20 t_0$ with~$t_0 =L/U_0$. We use a uniform Cartesian grid for all computations related to this problem with~$N_{\mathrm{ele}}=N_{\mathrm{ele},1\mathrm{D}}^d=(N_{\mathrm{ele},1\mathrm{D},l=0} 2^l)^d$ elements, where~$N_{\mathrm{ele},1\mathrm{D},l=0}$ is the number of elements on the coarsest mesh in one spatial dimension ($N_{\mathrm{ele},1\mathrm{D},l=0}=1$ unless specified otherwise) and~$l$ the refinement level. To compare results between different discretization approaches and polynomial degrees, it is appropriate to introduce an effective mesh resolution, which is defined as the number of nodes~$((k+1) N_{\mathrm{ele},1\mathrm{D}})^d$ for the present high-order nodal DG approach. In case of the compressible Navier--Stokes equation, the number of unknowns is~$d+2$ times the effective resolution. As mentioned above, we use~$n_{q}=\lceil \frac{3k+1}{2}\rceil$ quadrature points for all simulations unless otherwise specified.

\subsubsection{Selection of time step size}
To prepare the following performance measurements, the critical~$\mathrm{Cr}$ number and the critical~$\mathrm{D}$ number are determined. More importantly, the exponents~$e$ and~$f$ modeling the dependency on the polynomial degree~$k$ are verified to allow a fair comparison of the overall computational efficiency for different polynomial degrees. For this experiment, we consider polynomial degrees~$k=2,3,5,7,11,15$ with the corresponding number of cells~$20^3,16^3,10^3,8^3,5^3,4^3$, resulting in an effective resolution of either~$60^3$ or~$64^3$. To determine the critical~$\mathrm{Cr}$ number, a convection-dominated problem with large Reynolds number, here~$\mathrm{Re}=1600$, is considered. However, a viscous dominated problem for which the viscous time step restriction is the limiting factor, here~$\mathrm{Re}=1$, is used to determine the critical~$\mathrm{D}$ number. For the~$\mathrm{Re}=1$ test case, we reduce the end time of the simulation to~$t_{\mathrm{f}}=t_0$ in order to reduce costs. Based on our experience, we assume exponents of~$e=1.5$ and~$f=3.0$ in equation~\eqref{TimeStepRestrictions_Compressible} and verify this assumption numerically. In Table~\ref{CriticalCourantNumbers}, values of~$\mathrm{Cr}$ and~$\mathrm{D}$ are listed for which the simulation becomes unstable as well as slightly smaller values for which the simulation successfully completed until the final time~$t_{\mathrm{f}}$.
\begin{table}
\caption{Experimental determination of critical~$\mathrm{Cr}$ number and critical~$\mathrm{D}$ number as a function of the polynomial degree~$k$ using exponents of~$e=1.5$ and~$f=3.0$ in equation~\eqref{TimeStepRestrictions_Compressible}. The number of cells is chosen such that the effective resolution is either~$60^3$ or~$64^3$, i.e.,~$N_{\mathrm{ele},1\mathrm{D},l=0}=1$ for~$k=3,7,15$ and~$N_{\mathrm{ele},1\mathrm{D},l=0}=5$ for~$k=2,5,11$. For completeness, the critical Courant numbers~$\mathrm{Cr}_{\mathrm{inc}}$ specified in~\cite{Fehn2018b} for the incompressible Navier--Stokes solver are also listed.}
\label{CriticalCourantNumbers}
\renewcommand{\arraystretch}{1.1}
\begin{center}
\begin{tabular}{lllllll}
\hline
  & \multicolumn{6}{l}{Polynomial degree}\\
\cline{2-7}
					   & $k=2$ & $k=3$ & $k=5$ & $k=7$ & $k=11$ & $k=15$\\
\hline
$\mathrm{Cr}_{\mathrm{comp}} $ stable   & $0.68$ & $0.74$ & $0.76$ & $0.72$ & $0.64$ & $0.58$\\
$\mathrm{Cr}_{\mathrm{comp}} $ unstable & $0.70$ & $0.76$ & $0.78$ & $0.74$ & $0.66$ & $0.60$\\
\hline
$\mathrm{D}$ stable    & $0.032$ & $0.036$ & $0.036$ & $0.034$ & $0.026$ & $0.022$\\
$\mathrm{D}$ unstable  & $0.034$ & $0.038$ & $0.038$ & $0.036$ & $0.028$ & $0.024$\\
\hline
$\mathrm{Cr}_{\mathrm{inc}}$ stable   & $0.20$ & $0.21$ & $0.22$ & $0.19$ & $0.22$ & $0.18$\\
$\mathrm{Cr}_{\mathrm{inc}}$ unstable & $0.21$ & $0.22$ & $0.23$ & $0.20$ & $0.23$ & $0.19$\\
\hline
\end{tabular}
\end{center}
\renewcommand{\arraystretch}{1}
\end{table}
The fact that the critical~$\mathrm{Cr}$ and~$\mathrm{D}$ numbers are almost independent of the polynomial degree confirm our assumption regarding the choice of the exponents~$e, f$. Hence, all the following computations will be performed with~$e=1.5$ and~$f=3.0$, allowing to use constant~$\mathrm{Cr}$ and~$\mathrm{D}$ numbers over a wide range of polynomial degrees. Note that an exponent of~$1.5$ for the convective time step restriction has been found to be appropriate also for the incompressible Navier--Stokes solver in~\cite{Fehn2018b}. The following simulations for the Taylor--Green problem are simulated using~$\mathrm{Cr}=0.5$ and~$\mathrm{D}=0.02$ to ensure stability for all spatial resolutions. For comparison,~$\mathrm{Cr}_{\mathrm{inc}}=0.125$ has been used in~\cite{Fehn2018b} for all polynomial degrees. Hence the ratio between critical Courant number and the Courant number selected for the simulations is approximately the same for the compressible solver and the incompressible solver, allowing a fair comparison between both approaches in terms of computational costs. One can easily verify that the convective time step restriction is the limiting one for all spatial resolutions considered here. For example, the break-even point at which the viscous time step restriction becomes the limiting one would be refine level~$l=11$ for~$k=2$,~$l=10$ for~$k=3$,~$l=8$ for~$k=7$, and~$l=7$ for~$k=15$ corresponding to effective resolutions of~$6144^3$,~$4096^3$,~$2048^3$, and~$2048^3$, respectively. Evaluating equation~\eqref{RatioTimeStepSizesCompInc} with~$\mathrm{Cr}_{\mathrm{comp}}=0.5$,~$s=7$,~$\mathrm{M}=0.1$, and~$\mathrm{Cr}_{\mathrm{inc}}=0.125$ used for the incompressible DG solver yields~$\Delta t_{\mathrm{inc}} = 19.25 \cdot\Delta t_{s,\mathrm{comp}}$. While a significantly larger time step size is a potential performance advantage of the incompressible formulation, it is unclear whether this advantage of the incompressible solver pays off in terms of overall computational costs. This aspect will be analyzed in the following.

\begin{figure}[!ht]
 \centering 
 \subfigure[Effective resolution of~$64^3$ for different combinations of refinement level~$l$ and polynomial degree~$k$.]{
	\includegraphics[width=1.0\textwidth]{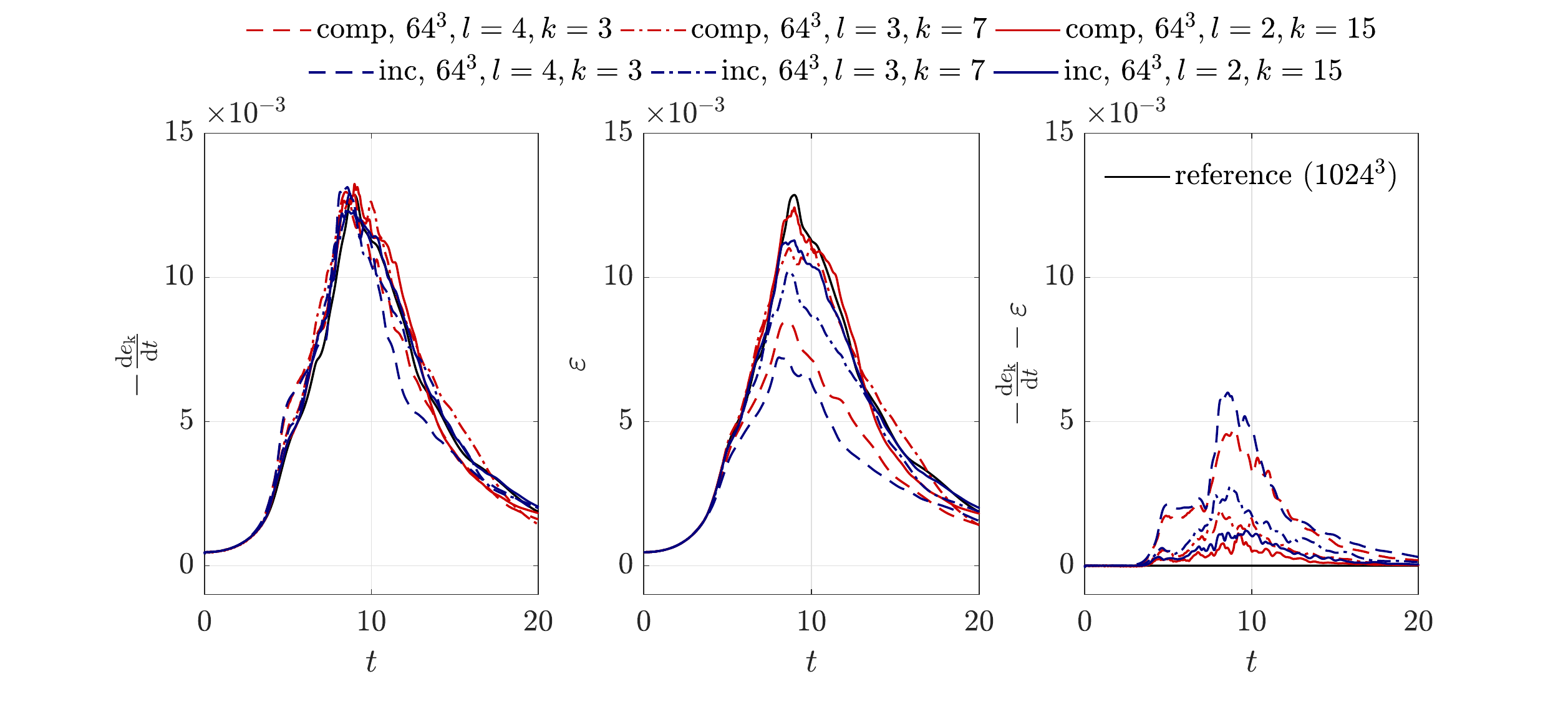}}
 \subfigure[Effective resolution of~$128^3$ for different combinations of refinement level~$l$ and polynomial degree~$k$.]{
	\includegraphics[width=1.0\textwidth]{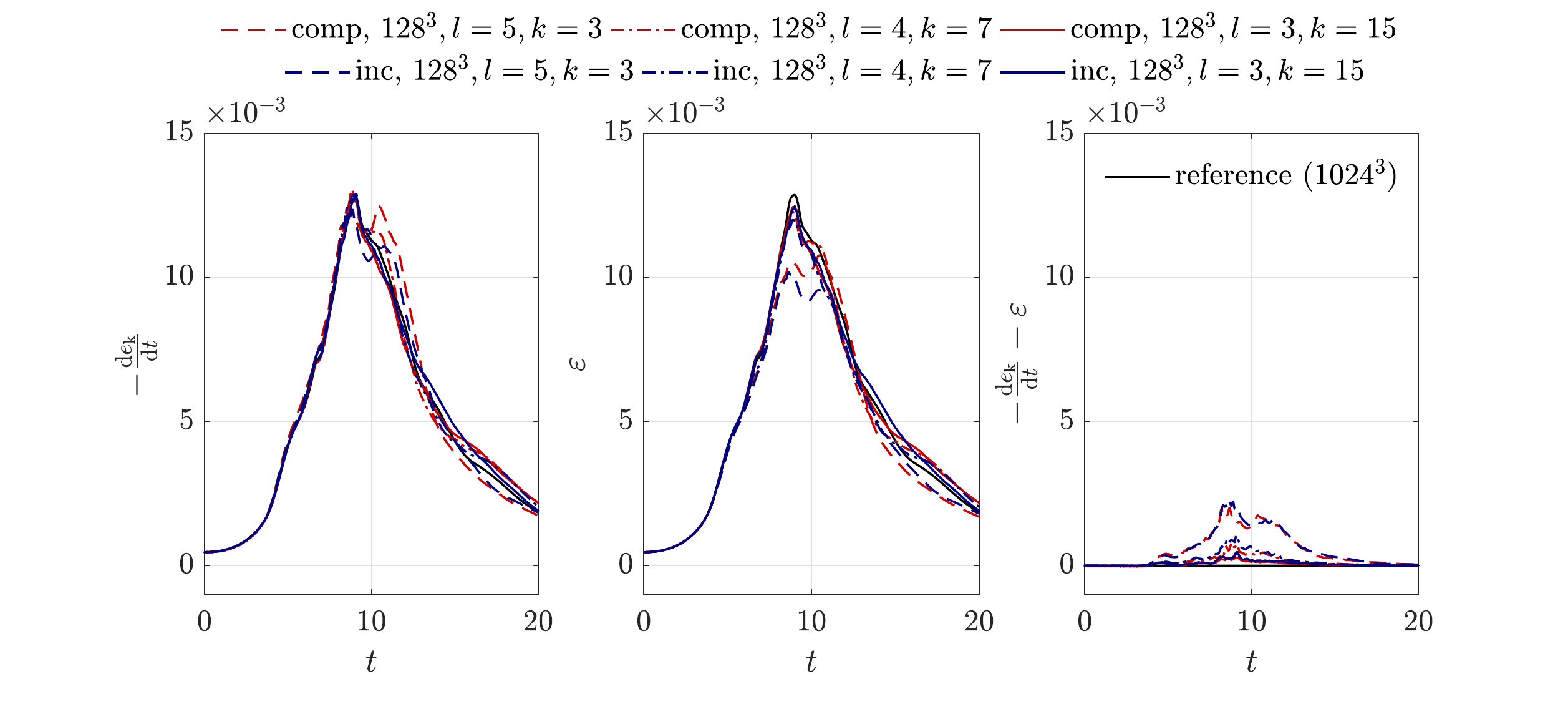}}
\caption{3D Taylor--Green vortex problem at~$\mathrm{Re}=1600$: Assessment of accuracy in terms of rate of change of kinetic energy, molecular dissipation, and numerical dissipation for effective resolutions of~$64^3$ and~$128^3$ and for polynomial degrees~$k=3,7,15$. Results for the incompressible flow solver published in~\cite{Fehn2018b} are also shown for the same spatial resolutions.}
\label{fig:3D_Taylor_Green_Convergence_Same_Effective_Resolution}
\end{figure}

\subsubsection{Accuracy of discretization scheme}
\begin{figure}[!ht]
 \centering 
\includegraphics[width=1.0\textwidth]{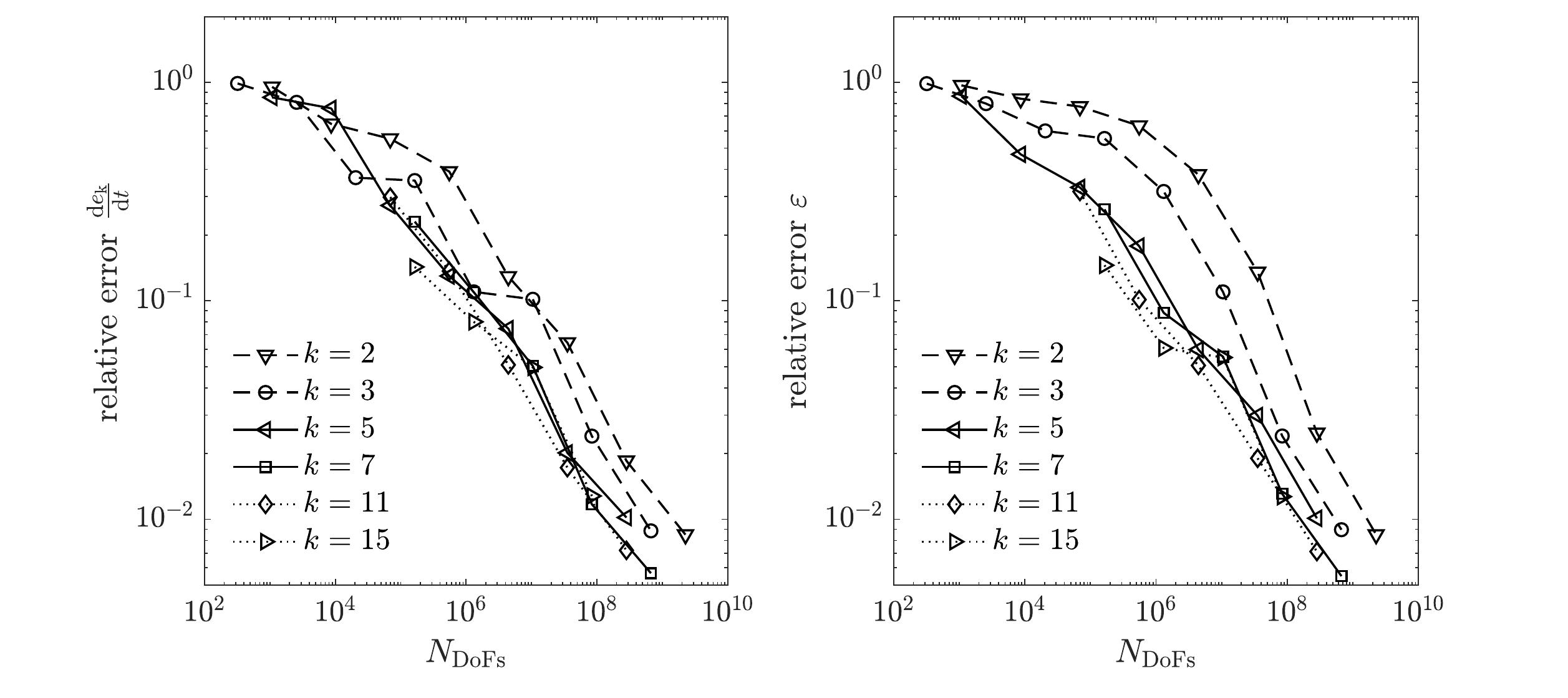}
\caption{3D Taylor--Green vortex problem at~$\mathrm{Re}=1600$: Efficiency of spatial discretization approach of compressible DG solver in terms of accuracy versus number of unknowns for polynomial degrees~$k=2,3,5,7,11,15$.}
\label{fig:3D_Taylor_Green_Efficiency_Spatial_Discretization}
\end{figure}
As usual, we consider the temporal evolution of the kinetic energy~$e_{\mathrm{k}}$, the kinetic energy dissipation rate~$\varepsilon$, and the numerical dissipation~$\varepsilon_{\mathrm{num}}$ to assess the accuracy of the numerical results. Since we use the compressible DG solver to compute incompressible flow problems, these quantities are calculated as for the incompressible case~\cite{Fehn18}
\begin{align*}
e_{\mathrm{k}} = \frac{1}{V_{\Omega_h}}\int_{\Omega_h} \frac{1}{2} \bm{u}_h\cdot \bm{u}_h\; \mathrm{d}\Omega \; ,\hspace{0,5cm}
\varepsilon = \frac{\nu}{V_{\Omega_h}} \int_{\Omega_h} \Grad{\bm{u}_h} : \Grad{\bm{u}_h}\; \mathrm{d}\Omega \; ,\hspace{0,5cm}
\varepsilon_{\mathrm{num}} = -\frac{\mathrm{d}e_{\mathrm{k}}}{\mathrm{d}t}-\varepsilon \; .
\end{align*}
Figure~\ref{fig:3D_Taylor_Green_Convergence_Same_Effective_Resolution} shows results for the temporal evolution of these quantities for polynomial degrees~$k=3,7,15$. The refinement level is chosen such that the effective resolution is the same for all polynomial degrees. To put the results for the compressible DG solver into perspective, we additionally plot the results for the incompressible DG solver published in~\cite{Fehn2018b} for a direct comparison of the accuracy of both solution strategies. For an effective resolution of~$64^3$, the compressible DG solver appears to be slightly more accurate than the incompressible solver exhibiting larger values for the dissipation rate~$\varepsilon$ slightly closer to the reference curve and a slightly reduced numerical dissipation as compared to the incompressible formulation. For an effective resolution of~$128^3$, no noticable differences can be observed between both formulations. Overall, both DG solvers predict the results accurately and convergence towards the (incompressible) reference solution. Moreover, we compute relative~$L^2$-errors of~$-\frac{\mathrm{d}e_{\mathrm{k}}(t)}{\mathrm{d}t}$ and~$\varepsilon(t)$ as defined in~\cite{Fehn18} for an objective and quantitative assessment of the accuracy of high-order discretizations. A detailed~$h$-convergence study for various polynomial degrees~$k=2,3,5,7,11,15$ is presented in Figure~\ref{fig:3D_Taylor_Green_Efficiency_Spatial_Discretization}, where the relative~$L^2$-errors are shown as a function of the number of unknowns. The results are in agreement with those published in~\cite{Fehn18,Fehn2018b} for the incompressible DG solver. The accuracy (efficiency of spatial discretization scheme) continuously improves for higher polynomial degrees~$k$. For very large polynomial degrees~$k\geq 5$, however, the differences are rather small and the accuracy of high-order methods somehow stagnates which can be explained by the under-resolution of the considered flow problem for the considered spatial resolutions, i.e., the relative errors are in the range of~$10^{-2}-10^{0}$ and we expect that asymptotic rates of convergence would only be observable for significantly finer DNS-like resolutions. Finally, let us note that a more relevant metric regarding the efficiency of high-order methods is the overall efficiency of the method in terms of accuracy over computational costs which will be analyzed in Section~\ref{TGV_OverallEfficieny}.

\begin{remark}
For high polynomial degrees~$k=7,11,15$ we detected instabilities for the compressible DG solver on the coarest possible mesh with only one cell. These simulations blew up even for very small~$\mathrm{Cr}$ numbers or when using~$n_q=2k+1$ quadrature points as dealiasing strategy to account for the cubic nonlinearities of the equations. For~$k=7$ and~$l=1$, the simulation remains stable but we detected significant negative values of the numerical dissipation. For example, this particular resolution crashed for a different Mach number,~$\mathrm{M}=0.2$. Note that a similar lack of robustness as detected here has also been reported in~\cite{Winters2017,Moura2017} for high-order DG discretizations of the compressible Navier--Stokes equations in the sense that consistent over-integration does not prevent instabilities for all meshes and Reynolds numbers. This is an aspect that could be addressed as part of future work.
\end{remark}

\subsubsection{Computational costs and overall efficiency}\label{TGV_OverallEfficieny}
\begin{figure}[!ht]
 \centering 
\includegraphics[width=1.0\textwidth]{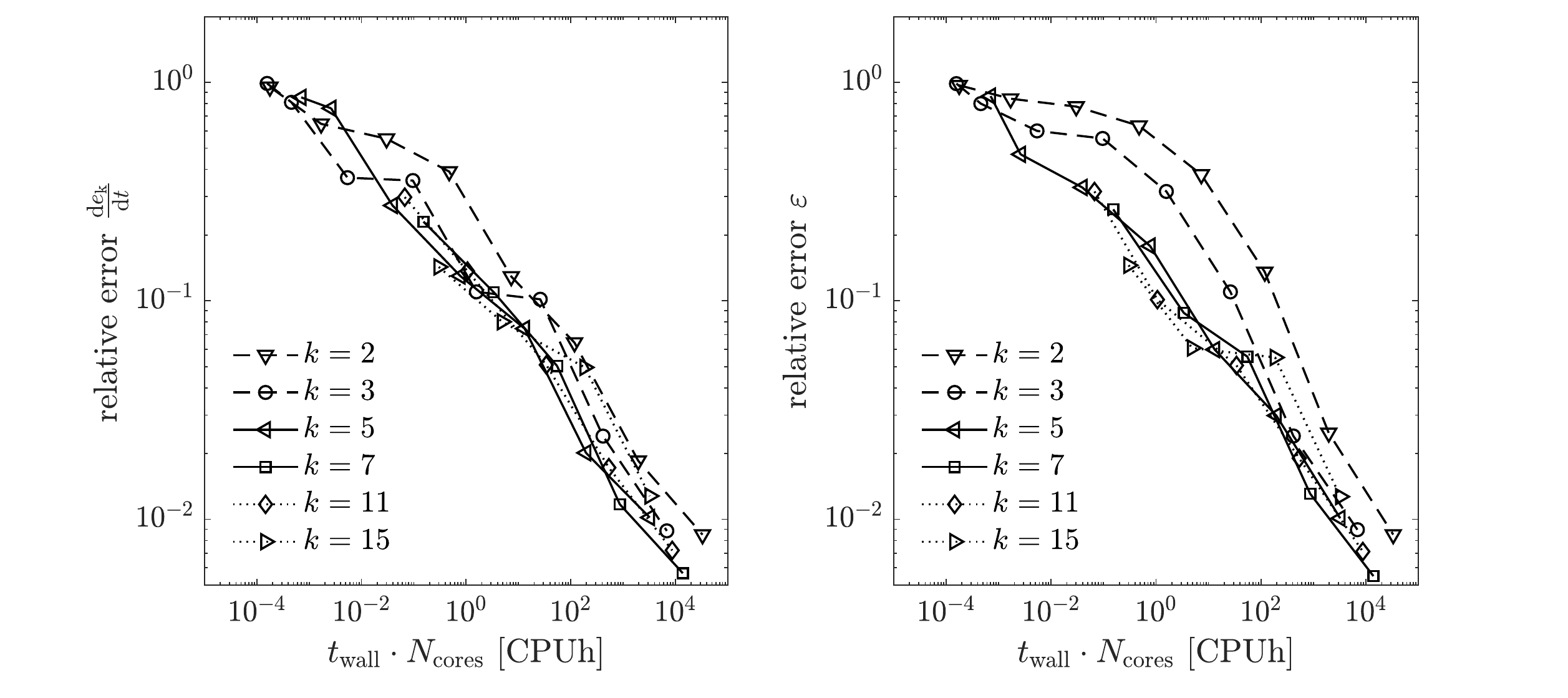}
\caption{3D Taylor--Green vortex problem at~$\mathrm{Re}=1600$: Overall efficiency of compressible DG solver in terms of accuracy versus computational costs for polynomial degrees~$k=2,3,5,7,11,15$.}
\label{fig:3D_Taylor_Green_Overall_Efficiency}
\end{figure}

As mentioned above, we are finally interested in the overall efficiency of the proposed compressible DG solver. Figure~\ref{fig:3D_Taylor_Green_Overall_Efficiency} shows the efficiency of the compressible DG solver in terms of accuracy versus computational costs for polynomial degrees~$k=2,3,5,7,11,15$.
The results for polynomial degrees~$k=5,7$ significantly improve the overall efficiency as compared to~$k=2,3$ and achieve the best overall efficiency. For very high-order methods with~$k=11,15$ the efficiency can not be further improved compared to~$k=5,7$ as has been observed in a similar manner in~\cite{Fehn2018b} for the incompressible formulation. Instead, the efficiency is slightly worse for very high-order methods which is mainly due to the time step restriction of the convective term according to the~$k^{1.5}$ relation on the one hand and the decrease in efficiency of the matrix-free implementation for very large polynomial degrees on the other hand. These effects render high-order methods~$k=11,15$ less efficient than methods with moderately large polynomial degree~$k=5,7$ for this under-resolved turbulent flow. As a conclusion, a significant improvement of very high-order methods in terms of the accuracy of the discretization scheme is a necessary prerequisite for high-order methods to be more efficient. An efficient matrix-free implementation with costs per degree of freedom almost independent of the polynomial degree~$k$ as demonstrated in Section~\ref{ResultsMatrixFreeImplementation} is also a key factor with respect to the efficiency of high-order methods (consider for example a matrix-free implementation with a theoretical complexity of~$\mathcal{O}(k)$ per degree of freedom or even a method relying (partially) on a matrix-based variant with complexity~$\mathcal{O}(k^{d})$ per degree of freedom and the impact this would have on the overall performance). For a more detailed discussion we refer to~\cite{Fehn2018b} where these aspects are thoroughly investigated for the incompressible DG solver.

To put the compressible DG solver into perspective, let us compare the overall computational costs to the incompessible formulation.
\begin{table}[!ht]
\caption{Performance results for polynomial degrees~$k=2,3,5,7,11,15$ and different mesh refinement levels~$l=3,...,8$. The time interval is~$0\leq t \leq t_{\mathrm{f}}$ with end time~$t_{\mathrm{f}}=20 t_0=20\frac{L}{U_0}$. The Courant number is~$\mathrm{Cr}=0.5$ for all computations. %(except for~$l=0$ and~$k=7$ where~$\mathrm{Cr}=0.25$ is used). 
The overall computational costs (including postprocessing) are compared to the incompressible solver in the last column that represents the speed-up of the incompressible solver against the compressible one, i.e., a value~$\geq 1$ indicates that the incompressible solver is faster.}
\label{tab:WallTimesAndCosts}
\renewcommand{\arraystretch}{1.1}
\begin{center}
\begin{tabular}{lllllllll}
\hline
$k$ & $l$ & resolution & $N_{\mathrm{DoFs}}$ & $N_{\Delta t}$ & $t_{\mathrm{wall}}$ [s] & $N_{\mathrm{cores}}$ & costs [CPUh] & speed-up (inc vs.~comp)\\
\hline
%2 & 1 & $6^3$    & $1.1\cdot 10^{3}$ & 397   & $6.4\cdot 10^{-1}$ & $1$                  & $1.8 \cdot 10^{-4}$ & 4.3\\
%  & 2 & $12^3$   & $8.6\cdot 10^{3}$ & 793   & $6.1\cdot 10^{0}$  & $1$                  & $1.7 \cdot 10^{-3}$ & 4.3\\
2 & 3 & $24^3$   & $6.9\cdot 10^{4}$ & 1585  & $1.4\cdot 10^{1}$  & $8$                  & $3.0 \cdot 10^{-2}$ & 2.3\\
  & 4 & $48^3$   & $5.5\cdot 10^{5}$ & 3170  & $6.1\cdot 10^{1}$  & $28$                 & $4.8 \cdot 10^{-1}$ & 2.1\\
  & 5 & $96^3$   & $4.4\cdot 10^{6}$ & 6339  & $9.4\cdot 10^{2}$  & $28$                 & $7.3 \cdot 10^{0}$  & 2.6\\
  & 6 & $192^3$  & $3.5\cdot 10^{7}$ & 12677 & $1.5\cdot 10^{4}$  & $28$                 & $1.2 \cdot 10^{2}$  & 2.2\\
  & 7 & $384^3$  & $3.5\cdot 10^{7}$ & 25353 & $3.1\cdot 10^{4}$  & $224$    			 & $2.0 \cdot 10^{3}$  & 2.1\\
  & 8 & $768^3$  & $2.3\cdot 10^{9}$ & 50706 & $6.6\cdot 10^{4}$  & $1792$ 			     & $3.3 \cdot 10^{4}$  & 2.1\\
\hline
%3 & 0 & $4^3$   & $3.2\cdot 10^{2}$ & 364   & $5.7\cdot 10^{-1}$ & $1$                  & $1.6 \cdot 10^{-4}$ & 3.1\\
%  & 1 & $8^3$   & $2.6\cdot 10^{3}$ & 728   & $1.7\cdot 10^{0}$  & $1$                  & $4.6 \cdot 10^{-4}$ & 2.3\\
%  & 2 & $16^3$  & $2.0\cdot 10^{4}$ & 1456  & $1.9\cdot 10^{1}$  & $1$                  & $5.4 \cdot 10^{-3}$ & 3.4\\
3 & 3 & $32^3$  & $1.6\cdot 10^{5}$ & 2912  & $4.3\cdot 10^{1}$  & $8$                  & $9.6 \cdot 10^{-2}$ & 2.2\\
  & 4 & $64^3$  & $1.3\cdot 10^{6}$ & 5823  & $2.0\cdot 10^{2}$  & $28$                 & $1.6 \cdot 10^{0}$  & 2.1\\
  & 5 & $128^3$ & $1.0\cdot 10^{7}$ & 11645 & $3.4\cdot 10^{3}$  & $28$                 & $2.6 \cdot 10^{1}$  & 2.0\\
  & 6 & $256^3$ & $8.4\cdot 10^{7}$ & 23289 & $2.7\cdot 10^{4}$  & $56$     			& $4.1 \cdot 10^{2}$  & 1.8\\
  & 7 & $512^3$ & $6.7\cdot 10^{8}$ & 46577 & $5.5\cdot 10^{4}$  & $448$   				& $6.9 \cdot 10^{3}$  & 1.8\\
\hline
%5 & 0 & $6^3$   & $1.1\cdot 10^{3}$ & 783   & $2.5\cdot 10^{0}$  & $1$                  & $7.1 \cdot 10^{-4}$ & 3.2\\
%  & 1 & $12^3$  & $8.6\cdot 10^{3}$ & 1566  & $9.5\cdot 10^{0}$  & $1$                  & $2.6 \cdot 10^{-3}$ & 2.2\\
%  & 2 & $24^3$  & $6.9\cdot 10^{4}$ & 3132  & $1.4\cdot 10^{2}$  & $1$                  & $3.9 \cdot 10^{-2}$ & 3.1\\
5 & 3 & $48^3$  & $5.5\cdot 10^{5}$ & 6264  & $3.5\cdot 10^{2}$  & $8$                  & $7.7 \cdot 10^{-1}$ & 2.4\\
  & 4 & $96^3$  & $4.4\cdot 10^{6}$ & 12528 & $1.7\cdot 10^{3}$  & $28$                 & $1.3 \cdot 10^{1}$  & 2.1\\
  & 5 & $192^3$ & $3.5\cdot 10^{7}$ & 25055 & $2.5\cdot 10^{4}$  & $28$                 & $2.0 \cdot 10^{2}$  & 1.7\\
  & 6 & $384^3$ & $2.8\cdot 10^{8}$ & 50109 & $5.2\cdot 10^{4}$  & $224$    			& $3.2 \cdot 10^{3}$  & 1.6\\
\hline
%7 & 0 & $8^3$   & $2.6\cdot 10^{3}$ & 2594  & $1.8\cdot 10^{1}$ & $1$ 			        & $4.9 \cdot 10^{-3}$ & 5.3\\
%  & 1 & $16^3$  & $2.0\cdot 10^{4}$ & 2594  & $3.6\cdot 10^{1}$ & $1$                   & $1.0 \cdot 10^{-2}$ & 1.6\\
%  & 2 & $32^3$  & $1.6\cdot 10^{5}$ & 5188  & $5.5\cdot 10^{2}$ & $1$                   & $1.5 \cdot 10^{-1}$ & 2.4\\
7 & 3 & $64^3$  & $1.3\cdot 10^{6}$ & 10376 & $1.5\cdot 10^{3}$ & $8$                   & $3.4 \cdot 10^{0}$  & 1.9\\
  & 4 & $128^3$ & $1.0\cdot 10^{7}$ & 20752 & $7.1\cdot 10^{3}$ & $28$ 					& $5.5 \cdot 10^{1}$  & 1.6\\
  & 5 & $256^3$ & $8.4\cdot 10^{7}$ & 41503 & $2.8\cdot 10^{4}$ & $112$    				& $8.7 \cdot 10^{2}$  & 1.5\\
  & 6 & $512^3$ & $6.7\cdot 10^{8}$ & 83005 & $1.1\cdot 10^{5}$ & $448$   				& $1.4 \cdot 10^{4}$  & 1.5\\
\hline
%11& 1 & $24^3$  & $8.6\cdot 10^{3}$ & 5110  & $2.4\cdot 10^{2}$ & $1$                   & $6.7 \cdot 10^{-2}$ & 0.83\\
%  & 2 & $48^3$  & $6.9\cdot 10^{4}$ & 10220 & $3.8\cdot 10^{3}$ & $1$                   & $1.1 \cdot 10^{0}$  & 1.7\\
11& 3 & $96^3$  & $5.5\cdot 10^{5}$ & 20439 & $1.5\cdot 10^{4}$ & $8$                   & $3.4 \cdot 10^{1}$  & 1.8\\
  & 4 & $192^3$ & $4.4\cdot 10^{6}$ & 40878 & $6.9\cdot 10^{4}$ & $28$ 					& $5.4 \cdot 10^{2}$  & 1.6\\
  & 5 & $384^3$ & $3.5\cdot 10^{7}$ & 81755 & $1.4\cdot 10^{5}$ & $224 $    			& $8.7 \cdot 10^{3}$  & 1.6\\
\hline
%15& 1 & $32^3$  & $1.6\cdot 10^{5}$ & 8137  & $1.1\cdot 10^{3}$ & $1$  			        & $3.0 \cdot 10^{-1}$ & 0.521\\
15& 2 & $64^3$  & $1.3\cdot 10^{6}$ & 16274 & $1.8\cdot 10^{4}$ & $1$                   & $4.9 \cdot 10^{0}$  & 1.3\\
  & 3 & $128^3$ & $1.0\cdot 10^{7}$ & 32547 & $8.4\cdot 10^{4}$ & $8$                   & $1.9 \cdot 10^{2}$  & 1.5\\
  & 4 & $256^3$ & $8.4\cdot 10^{7}$ & 65093 & $1.4\cdot 10^{5}$ & $84$ 				    & $3.2 \cdot 10^{3}$  & 1.4\\
\hline
\end{tabular}
\end{center}
\renewcommand{\arraystretch}{1}
\end{table}
\begin{table}[!h]
\caption{Sensitivity of computational costs with respect to numerical solver parameters such as the chosen relative solver tolerance \texttt{reltol} for the incompressible flow solver and the chosen Mach number~$\mathrm{M}$ for the compressible flow solver. Computational costs in CPUh are specified for polynomial degree~$k=7$ and refinement levels~$l=2,...,6$ for different values of the parameters.}\label{tab:3D_Taylor_Green_Vortex_Sensitivity_Parameters}
\renewcommand{\arraystretch}{1.1}
\begin{center}
\begin{tabular}{ccccccccc}
\hline
& &\multicolumn{3}{c}{Costs for incompressible solver [CPUh]} & &\multicolumn{3}{c}{Cost for compressible solver [CPUh]}\\
\cline{3-5} \cline{7-9}
$k$ & $l$ & \texttt{reltol=1e-2} & \texttt{reltol=1e-3} & \texttt{reltol=1e-6} & & $\mathrm{M}=0.3$ & $\mathrm{M}=0.2$ & $\mathrm{M}=0.1$\\
\hline
7 & 2 & $2.8 \cdot 10^{-2}$ & $3.9 \cdot 10^{-2}$ & $6.4 \cdot 10^{-2}$ & & $6.0 \cdot 10^{-2}$ & $8.2 \cdot 10^{-2}$ & $1.5 \cdot 10^{-1}$\\
  & 3 & $8.0 \cdot 10^{-1}$ & $1.0 \cdot 10^{0}$  & $1.8 \cdot 10^{0}$  & & $1.3 \cdot 10^{0}$  & $1.9 \cdot 10^{0}$  & $3.4 \cdot 10^{0}$\\
  & 4 & $1.5 \cdot 10^{1}$  & $2.0 \cdot 10^{1}$  & $3.4 \cdot 10^{1}$  & & $2.2 \cdot 10^{1}$  & $3.0 \cdot 10^{1}$  & $5.5 \cdot 10^{1}$\\
  & 5 & $2.6 \cdot 10^{2}$  & $3.3 \cdot 10^{2}$  & $5.9 \cdot 10^{2}$  & & $3.4 \cdot 10^{2}$  & $4.7 \cdot 10^{2}$  & $8.7 \cdot 10^{2}$\\
  & 6 & $4.0 \cdot 10^{3}$  & $5.3 \cdot 10^{3}$  & $9.4 \cdot 10^{3}$  & & $6.8 \cdot 10^{3}$  & $7.6 \cdot 10^{3}$  & $1.4 \cdot 10^{4}$\\
\hline
\end{tabular}
\end{center}
\renewcommand{\arraystretch}{1}
\end{table}
As shown above, the time step size for the incompressible solver is significantly larger than the time step per Runge--Kutta stage for the compressible solver. However, we recall that three linear systems of equations (pressure Poisson, projection, and Helmholtz-like equations) have to be solved for the incompressible solver, while one only has to evaluate the nonlinear Navier--Stokes operator per~$\Delta t_{s,\mathrm{comp}}$ in case of the compressible solver. At the same time, the matrix-free operator evaluation can be performed at a higher speed for the incompressible solver due to simpler operators and due to over-integration reducing the efficiency of the compressible solver, see Section~\ref{ResultsMatrixFreeImplementation}. In Table~\ref{tab:WallTimesAndCosts}, we list the computational costs of the compressible DG solver for various spatial resolutions and compare the overall computational costs to the incompressible formulation~\cite{Fehn2018b} in the last column. The simulations have been performed on the same hardware (Intel Xeon CPU E5-2697 v3) for both solvers allowing a direct comparison of computational costs. For lower polynomial degrees, the incompressible solver is approximately a factor of~$2$ faster than the compressible solver and a factor of~$1.5$ towards very high polynomial degrees. This result is interesting with respect to several aspects: Although the compressible and incompressible formulations differ significantly regarding the number of time steps and the computational costs per time step, the two effects seem to counterbalance each other so that both approaches allow to solve the problem with a similar amount of overall computational costs. Recalling that~$\Delta t_{\mathrm{inc}}$ is approximately~$20$ times larger than~$\Delta t_{s,\mathrm{comp}}$, this implies that solving one time step for the incompressible solver is approximately~$10-13$ times more expensive than evaluating one Runge--Kutta stage for the compressible solver. One might argue that these numbers point to a significant advantage of the incompressible formulation but one should keep in mind that there are numerical solver parameters that have a significant impact on the overall performance of both solvers. These parameters are mainly the solver tolerances selected for the solution of linear systems of equations in case of the incompressible solver and the selected Mach number in case of the compressible solver. For example, it is often argued that the compressible formulation can not be competitive due to the severe time step restrictions in case of small~$\mathrm{M}$ numbers. Since we use a compressible formulation to solve incompressible flow problems, the Mach number can be considered as a numerical solver parameter rather than a physical quantity as long as the Mach number is small enough and does not deteriorate the accuracy of the results. In order to demonstrate the sensitivity of the overall computational costs with respect to these parameters, we present in Table~\ref{tab:3D_Taylor_Green_Vortex_Sensitivity_Parameters} results for different values of the relative solver tolerances~\footnote{The term ``relative solver tolerance'' means that the residual of the linear system of equations is reduced by a factor of~\texttt{reltol} compared to the initial residual where the initial guess of the solution is based on a second order extrapolation of the solution from previous instants of time.} (for the incompressible solver) and the Mach number (for the compressible solver). We repeated the simulations for the most efficient polynomial degree~$k=7$ for larger solver tolerances of~\texttt{reltol=1e-2} and~\texttt{reltol=1e-3} (compared to \texttt{reltol=1e-6}) and for larger Mach numbers of~$\mathrm{M}=0.3$ and~$\mathrm{M}=0.2$ (compared to~$\mathrm{M}=0.1$). For~$\mathrm{M}=0.3$ a slightly smaller Courant number of~$\mathrm{Cr}=0.4$ had to be used for the finest mesh with~$l=6$ to obtain stability. Under idealized assumption one would expect a reduction of computational costs by a factor of three or two. Our numerical results in fact reveal the large impact of these parameters on computational costs. Depending on the solver parameters, there is now an overlap in computational costs between the compressible formulation and the incompressible formulation as can be seen from Table~\ref{tab:3D_Taylor_Green_Vortex_Sensitivity_Parameters}. At the same time, it has to be verified that changing the solver parameters does not deteriorate the accuracy of the results.
\begin{figure}[!ht]
 \centering 
\includegraphics[width=1.0\textwidth]{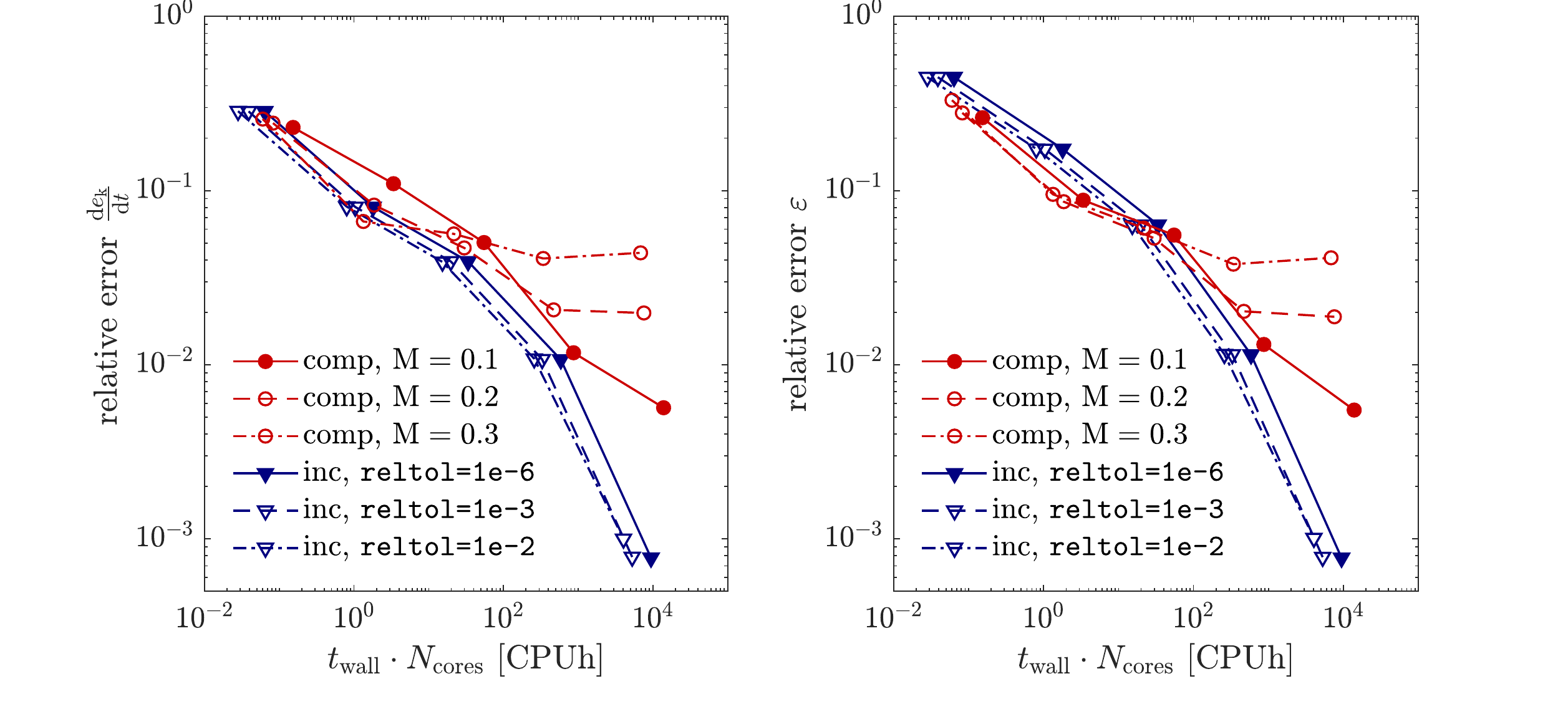}
\caption{3D Taylor--Green vortex problem at~$\mathrm{Re}=1600$: Comparison of efficiency of compressible DG solver to incompressible DG solver and sensitivity of solver parameters.}
\label{fig:3D_Taylor_Green_Sensitivity_Parameters}
\end{figure}
Figure~\ref{fig:3D_Taylor_Green_Sensitivity_Parameters} shows an~$h$-convergence plot for~$k=7$ for different values of the solver parameters. While the incompressible solver is more efficient with respect to the rate of change of the kinetic energy on coarse meshes, the compressible solver is more efficient regarding the dissipation rate~$\varepsilon$ on coarse meshes. These results also demonstrate that a small Mach number is necessary to ensure that the compressible solver converges to the same reference solution as the incompressible solver. For larger Mach numbers, the convergence of the compressible solver levels off due to the modelling error that becomes dominant over the discretization error. Hence, a compromise between accuracy and computational efficiency has to be found for the compressible Navier--Stokes solver with respect to the choice of the parameter~$\mathrm{M}$. For the incompressible solver, however, increasing the relative solver tolerance has almost no impact on the accuracy of the results. These results show that the value of~\texttt{reltol=1e-6} used in~\cite{Fehn2018b} appears to be too conservative and that the costs for the incompressible solver can be reduced by at least a factor of two without impacting the accuracy of the results. Reconsidering the results in Table~\ref{tab:WallTimesAndCosts}, we note that a solver tolerance of~\texttt{reltol=1e-3} would result in a speed-up by a factor of~$2-4$ for the incompressible solver. We conclude that the compressible solver can be competitive for this flow problem (as it is also a matter of solver parameters which approach is computationally cheaper) but that we also observe an advantage for the incompressible solver in terms of overall runtime.

Compared to results published recently in the literature for high-order DG discretizations of the compressible Navier--Stokes equations applied to the 3D Taylor--Green vortex problem, our approach achieves a significant improvement in computational costs. Compared to the results in~\cite{Wiart14} for a~$256^3$ resolution with polynomial degree~$k=3$, where costs of~$6.4\cdot 10^3$ CPUh are specified for~$t_{\mathrm{f}}=10$, our approach is approximately~$30$ times faster, see costs of~$4.1\cdot 10^2$ CPUh for this specific resolution and end time~$t_{\mathrm{f}}=20$ in Table~\ref{tab:WallTimesAndCosts}. Using~$l=2$ and~$k=15$ along with~$2k+1$ quadrature points to reproduce the simulations in~\cite{Gassner2013} with computational costs of~$83.2$ CPUh for~$t_{\mathrm{f}}=10$, the present approach requires costs of~$12.9$ CPUh for~$t_{\mathrm{f}}=20$, hence achieving a speedup by a factor of approximately~$13$. One should of course take into account that the results from literature have been obtained on different computer hardware that might be a factor of~$2-4$ slower compared to the one used in the present study. Nevertheless, these results are encouraging in the sense that many high-order DG implementations offer great potential for further performance improvements by means of code optimization, see also the discussion of results presented in Table~\ref{tab:Performance_Index}.

\subsection{Turbulent channel flow}
In the previous section, we compared the performance of both solvers for a geometrically simple problem on a uniform Cartesian mesh with cells of aspect ratio~$1$. An interesting question remains whether the situation regarding the convective and viscous time step restrictions changes on more complex meshes with large aspect ratios such as wall-bounded flows for which the mesh is typically refined towards the walls in order to resolve steep gradients occurring for large Reynolds numbers. The turbulent channel flow problem is an ideal test case for this investigation since it is a wall-bounded flow and at the same time one of the few turbulent test cases that is well-suited for a quantitative assessment of results.

\subsubsection{Problem description}

The turbulent channel flow problem is solved on a rectangular domain with dimensions~$(L_1,L_2,L_3)=(2\pi \delta, 2 \delta, \pi \delta)$ where periodic boundary conditions are applied in streamwise~($x_1$) and spanwise~($x_3$) directions and no-slip boundary conditions at the channels walls~($x_2=\pm \delta$). The flow is driven by a constant body force~$f_1$ in~$x_1$-direction. With the friction Reynolds number defined as~$\mathrm{Re}_{\tau}=u_{\tau} \delta /\nu$ and the wall friction velocity~$u_{\tau}=\sqrt{\tau_{\mathrm{w}}/\rho}$, a balance of forces in~$x_1$-direction yields~$f_1=\rho u_{\tau}^2/\delta$. We use the parameters~$\delta=1,f_1=\rho=1$ as in~\cite{Fehn18}, which implies~$u_{\tau}=1$, and~$\nu=1/\mathrm{Re}_{\tau}$. As in the previous section we use~$\mathrm{Pr}={\mu c_p}/{\lambda}=0.71$,~$\gamma={c_p}/{c_v}=1.4$, and~$\mathrm{R}=287$. The Mach number is set to~$\mathrm{M}={\Vert \bm{u} \Vert_{\mathrm{max}}}/{c}=0.1$ to mimic incompressible flows, where the mean centerline velocity is used as an estimation of the maximum velocity,~$\Vert \bm{u} \Vert_{\mathrm{max}}=\left< u_1\right>(x_2=0)$. This velocity is known from the DNS reference results (e.g.,~$\left< u_1\right>(x_2=0)=18.3$ for~$\mathrm{Re}_{\tau}=180$ and~$\left< u_1\right>(x_2=0)=22.4$ for~$\mathrm{Re}_{\tau}=950$) and is used for the estimation of the convective time step restrictions in equations~\eqref{TimeStepRestrictions_Compressible} and~\eqref{CFL_Condition}. A uniform density field~$\rho(\bm{x},t=0)=\rho_0=1$ is prescribed as initial condition as well as a uniform temperature field~$T(\bm{x},t=0)=T_0= c^2/(\gamma R)$ with~$c=\Vert \bm{u} \Vert_{\mathrm{max}}/\mathrm{M}$. At the channel walls, homogeneous Neumann boundary conditions are prescribed for the density and homogeneous Dirichlet boundary conditions for the temperature,~$T(x_2 = \pm \delta) = T_0$. The velocity field is initialized with the mean flow~$u_1=\Vert \bm{u} \Vert_{\mathrm{max}}\left(1-(x_2/\delta)^6\right)$ and additional small perturbations to initiate a turbulent flow state at~$t=0$. As in the previous example we use a Cartesian grid with refinement level~$l=0$ and~$N_{\mathrm{ele},1\mathrm{D},l=0}=1$ elements per coordinate direction on the coarsest level~$l=0$. As already indicated, the mesh is refined towards the no-slip walls according to the function~$f: [0,1]\rightarrow [-\delta,\delta]$
\begin{align}
x_2 \mapsto f(x_2) = \delta \frac{\tanh(C(2x_2-1))}{\tanh(C)} \; ,
\end{align}
which has been used for example in~\cite{Krank2017,Fehn18}. In the present work, we use a grid stretch factor of~$C = 1.8$ for all simulations and use an isoparametric mapping of degree~$k$. As illustrated in~\cite{Fehn2018b} for various refinement levels, this mesh deformation results in thin element layers with a high aspect ratio close to the walls. The maximum aspect ratio is~$h_{x_1}/h_{x_2} = 3.1$ for refine level~$l=1$,~$h_{x_1}/h_{x_2}  = 6.5$ for~$l=2$,~$h_{x_1}/h_{x_2} = 10.2$ for~$l=3$, and~$h_{x_1}/h_{x_2} = 12.8$ for~$l=4$. 

We define the characteristic time scale~$t_0=L_1/\Vert \bm{u} \Vert_{\mathrm{max}}$ as the flow-through time based on the mean centerline velocity. The simulations are performed over a time intervall of~$t_{\mathrm{f}}=200 t_0$, where the sampling of statistical results is performed over the time intervall~$100 t_0 \leq t \leq 200 t_0$. Statistical data is collected every~$10^{\mathrm{th}}$ time step.
% for the incompressible solver and every~$100^{\mathrm{th}}$ time step for the compressible solver due to smaller time step sizes for the latter solver. 
The statistical quantities considered in the following are the mean streamwise velocity~$u_1^+=\left< u_1 \right>/u_{\tau}$, the rms-velocities~$\left(u_i^{\prime}\right)^+ = \mathrm{rms}(u_i)/u_{\tau} = \sqrt{\left<{u_i^{\prime}}^2\right>}/u_{\tau}$, and the Reynolds shear stress~$(u_1^{\prime}u_2^{\prime})^+ = \left<u_1^{\prime}u_2^{\prime}\right>/u_{\tau}^2$. The dimensionless wall normal coordinate is given as~$x_2^+ = (x_2+1)/l^+$ where~$l^+=\nu/u_{\tau}$.

Based on the results in Section~\ref{TaylorGreenVortex}, a relative solver tolerance of~\texttt{reltol=1e-3} appears to be sufficient for the incompressible solver and will be used for the following simulations. The results in Section~\ref{TaylorGreenVortex} also revealed that a small Mach number has to be used for the compressible solver to enable an accurate solution close to the incompressible formulation. For this reason, we use~$\mathrm{M}=0.1$ for all computations. This value has for example also been used in~\cite{Wiart15} for a high-order compressible DG solver applied to turbulent channel flow and is the default value used in~\cite{Beck2014b} for various turbulent test cases.

\subsubsection{Time step restrictions}
Since the selection of the time step size has a large impact on the overall computational costs, we investigate the time step restrictions for both the compressible solver and the incompressible solver as a prerequisite for the performance measurements detailed below. On the one hand, it is unclear whether the convective time step restriction or the viscous time step restriction is the limiting one for the channel flow problem with stretched elements. On the other hand, we want to investigate whether the stretched mesh has an impact on the convective time step restriction for the compressible solver in relation to the incompressible solver. Since very high-order methods were found to be less efficient than moderate high-order methods for the Taylor--Green vortex problem in Section~\ref{TaylorGreenVortex}, we focus on polynomial degrees~$k\leq 7$ in this section.

% HIGH-ORDER MAPPING, GRID_STRETCH_FACTOR = 1.8
\begin{table}
\caption{Turbulent channel flow problem: Experimental determination of critical~$\mathrm{Cr}$ number and critical~$\mathrm{D}$ number as a function of the polynomial degree~$k$ using exponents of~$e=1.5$ and~$f=3.0$. The refinement level is~$l=3$ with~$N_{\mathrm{ele},1\mathrm{D},l=0}=1$ for all polynomial degrees~$k=2,3,5,7$.}
\label{tab:CriticalCourantNumbers_TurbulentChannel}
\renewcommand{\arraystretch}{1.1}
\begin{center}
\begin{tabular}{lllll}
\hline
& \multicolumn{4}{l}{Polynomial degree}\\
 \cline{2-5} 
& $k=2$ & $k=3$ & $k=5$ & $k=7$\\
\hline
$\mathrm{Cr}_{\mathrm{comp}} $ stable   & $1.5$ & $1.6$ & $1.5$ & $1.3$\\
$\mathrm{Cr}_{\mathrm{comp}} $ unstable & $1.6$ & $1.7$ & $1.6$ & $1.4$\\
\hline
$\mathrm{D}$ stable    & $0.15$ & $0.16$ & $0.15$ & $0.13$\\
$\mathrm{D}$ unstable  & $0.16$ & $0.17$ & $0.16$ & $0.14$\\
\hline
$\mathrm{Cr}_{\mathrm{inc}}$ stable   & $1.3$ & $1.7$ & $1.8$ & $1.8$\\
$\mathrm{Cr}_{\mathrm{inc}}$ unstable & $1.4$ & $1.8$ & $1.9$ & $1.9$\\
\hline
\end{tabular}
\end{center}
\renewcommand{\arraystretch}{1}
\end{table}
Table~\ref{tab:CriticalCourantNumbers_TurbulentChannel} shows results of a stability experiment in which the critical~$\mathrm{Cr}$ and~$\mathrm{D}$ numbers are determined. The polynomial degrees considered for this experiment are~$k=2,3,5,7$ for refinement level~$l=3$ (the corresponding effective mesh resolutions are~$24^3$,~$32^3$,~$48^3$, and~$64^3$) and the exponents are set to~$e=1.5$ and~$f=3.0$. To determine the critical~$\mathrm{Cr}$ numbers we consider the~$\mathrm{Re}_{\tau}=180$ test case. To obtain the critical~$\mathrm{D}$ values a viscous dominated problem with~$\nu = 1$ is used and the end time is set to~$t_{\mathrm{f}}=1$ for this laminar test case to reduce computational costs. For the compressible solver, the critical~$\mathrm{Cr}$ number is approximately a factor of~$10$ larger than the critical~$\mathrm{D}$ number as has been observed in a similar manner for the Taylor--Green example in Section~\ref{TaylorGreenVortex}. 
For the incompressible solver, the critical Courant number is as large as for the compressible solver. This differs from the results for the Taylor--Green vortex problem where~$\mathrm{Cr}_{\mathrm{inc}}$ has been noticably smaller than~$\mathrm{Cr}_{\mathrm{comp}}$. An explanation could be that the acoustic waves in the compressible case are isotropic in nature so that the minimum vertex distance is relevant for the time step restriction. Since the flow in~$x_1$-direction is aligned with the elements being stretched in streamwise direction, the relevant mesh size for the incompressible solver is larger than the minimum vertex distance in wall-normal direction. Since these element-local effects relevant for the incompressible solver can not be represented by the global Courant criterion~\eqref{CFL_Condition} selected in the present work for didactical reasons, the measured critical Courant numbers are problem dependent. To obtain problem independent critical Courant numbers, more sophisticated Courant criteria can be used, e.g., by evaluating the Courant condition element-wise taking into account the local velocity and the local mesh size in combination with adaptive time stepping~\cite[Appendix D]{Krank2018}. For example, using such an advanced criterion we obtain critical Courant numbers for the incompressible solver in the range~$0.16-0.19$ for~$\mathrm{Re}_{\tau}=180$ (and~$0.14-0.2$ for~$\mathrm{Re}_{\tau}=950$) for polynomial degrees~$k=2,3,5,7$. These numbers agree very well with those shown in Table~\ref{CriticalCourantNumbers} for a uniform mesh and the global Courant criterion. Since the focus of the present work is on the performance of the compressible solver relative to the incompressible solver, the global Courant criteria used here are well-suited to characterize the differences in efficiency between both solution approaches. As commented below, using adaptive time stepping techniques has only a small to moderate impact on the overall costs for this statistically steady problem and would not change the conclusions drawn in this work.

It turns out that the viscous time step restriction in case of the compressible solver is less of a concern for this specific problem and the applied mesh stretching despite the~$h^2$ and~$k^3$ dependency for the viscous time step limitation (as compared to~$h$ and~$k^{1.5}$ for the convective term). The reason is the small viscosity and the fact that the flow is under-resolved for all practical computations. The convective time step is one to two orders of magnitude smaller than the critical viscous time step for the spatial resolutions listed in Table~\ref{tab:CriticalCourantNumbers_TurbulentChannel}. This implies that the viscous time step restriction would only become relevant for significantly finer meshes resulting in highly-resolved DNS resolutions. The situation is the same for the LES simulations performed below for a larger Reynolds number of~$\mathrm{Re}_{\tau}=950$ and we expect that this holds true also for very large Reynolds numbers. To obtain accurate results for the turbulent channel flow problem a rule of thumb is that the node closest to the wall fulfills~$\Delta x_2^+ = \Delta x_2/l^+ \approx \mathrm{const} \approx 1$. This implies~$h_{\mathrm{min}}\sim l^+ \sim \nu$. Accordingly, in the limit~$\nu \rightarrow 0$, the viscous time step restriction depends linearly on~$h$ (or~$\nu$) according to equation~\eqref{TimeStepRestrictions_Compressible} just as the convective time step restriction.

\subsubsection{Accuracy of numerical results for~$\mathrm{Re}_{\tau}=180$ and~$\mathrm{Re}_{\tau}=950$}
To assess the accuracy of the present solvers for the turbulent channel flow problem, we perform a~$k$-refinement study for the two Reynolds numbers~$\mathrm{Re}_{\tau}=180$ and~$\mathrm{Re}_{\tau}=950$ and compare the results to accurate DNS reference data from~\cite{Moser99,delAlamo2004} denoted as DNS MKM99 and DNS AJZM04 in the following. An~$h$-refinement study for the same Reynolds numbers has been shown in~\cite{Fehn18} for the incompressible flow solver. Since every discretization scheme and every LES model has its limitations, in our opinion it is a necessity to explicitly demonstrate the limit in resolution at which the results start to deviate substantially from the reference solution.
We also mention that both the compressible and the incompressible DG solvers are generic and parameter-free turbulent flow solvers and that the same setup is used for all computations. For the lower Reynolds number of~$\mathrm{Re}_{\tau}=180$ we consider refinement level~$l=3$ with~$N_{\mathrm{ele},1\mathrm{D},l=0}=1$ and increasing polynomial degrees~$k=2,3,5,7$. The corresponding effective mesh resolutions are~$24^3$,~$32^3$,~$48^3$, and~$64^3$. For the higher Reynolds number of~$\mathrm{Re}_{\tau}=950$ the refinement level is set to~$l=4$, resulting in effective mesh resolutions of~$48^3$,~$64^3$,~$96^3$, and~$128^3$ for the different polynomial degrees~$k=2,3,5,7$.

% HIGH-ORDER MAPPING
\begin{figure}[!ht]
 \centering 
 \subfigure[Compressible solver.]{
	\includegraphics[width=1.0\textwidth]{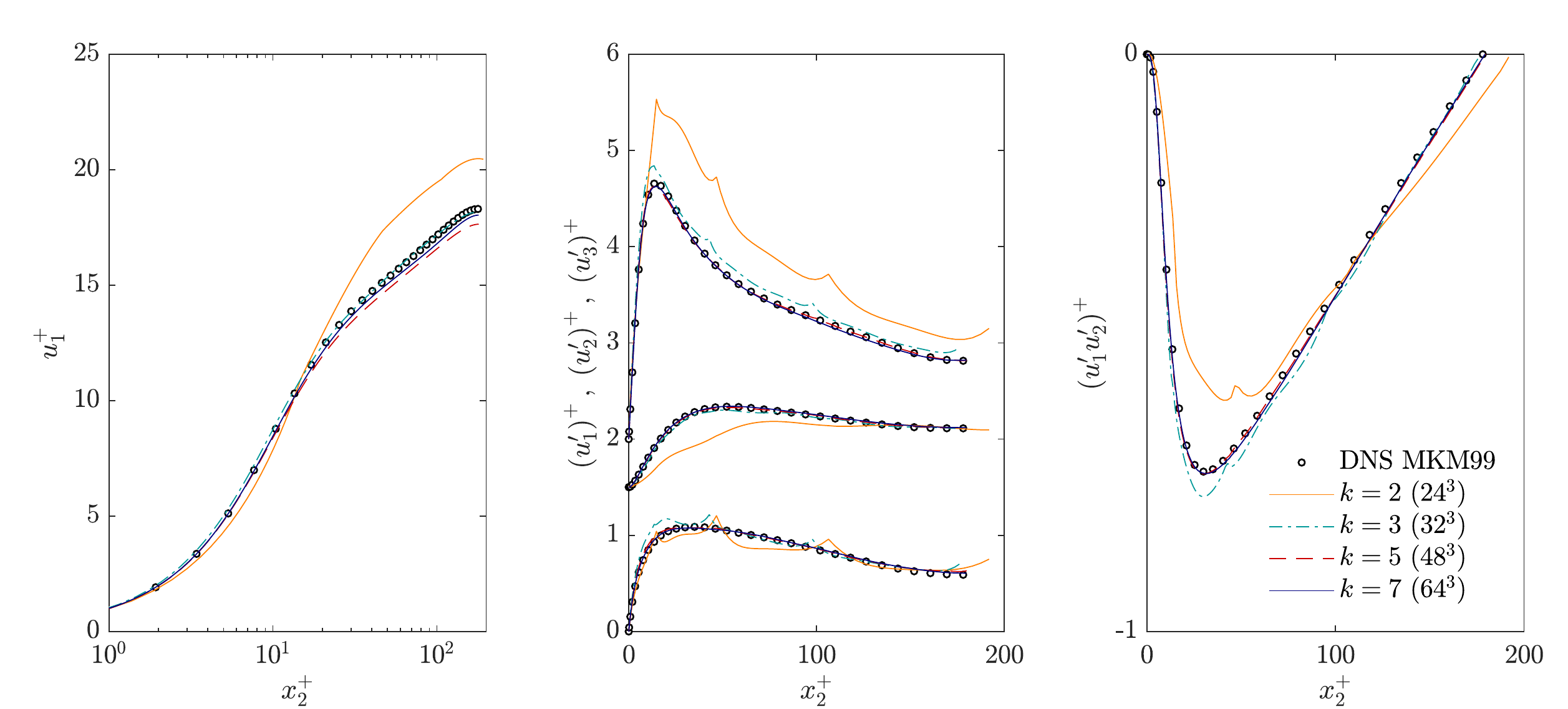}}
 \subfigure[Incompressible solver.]{
	\includegraphics[width=1.0\textwidth]{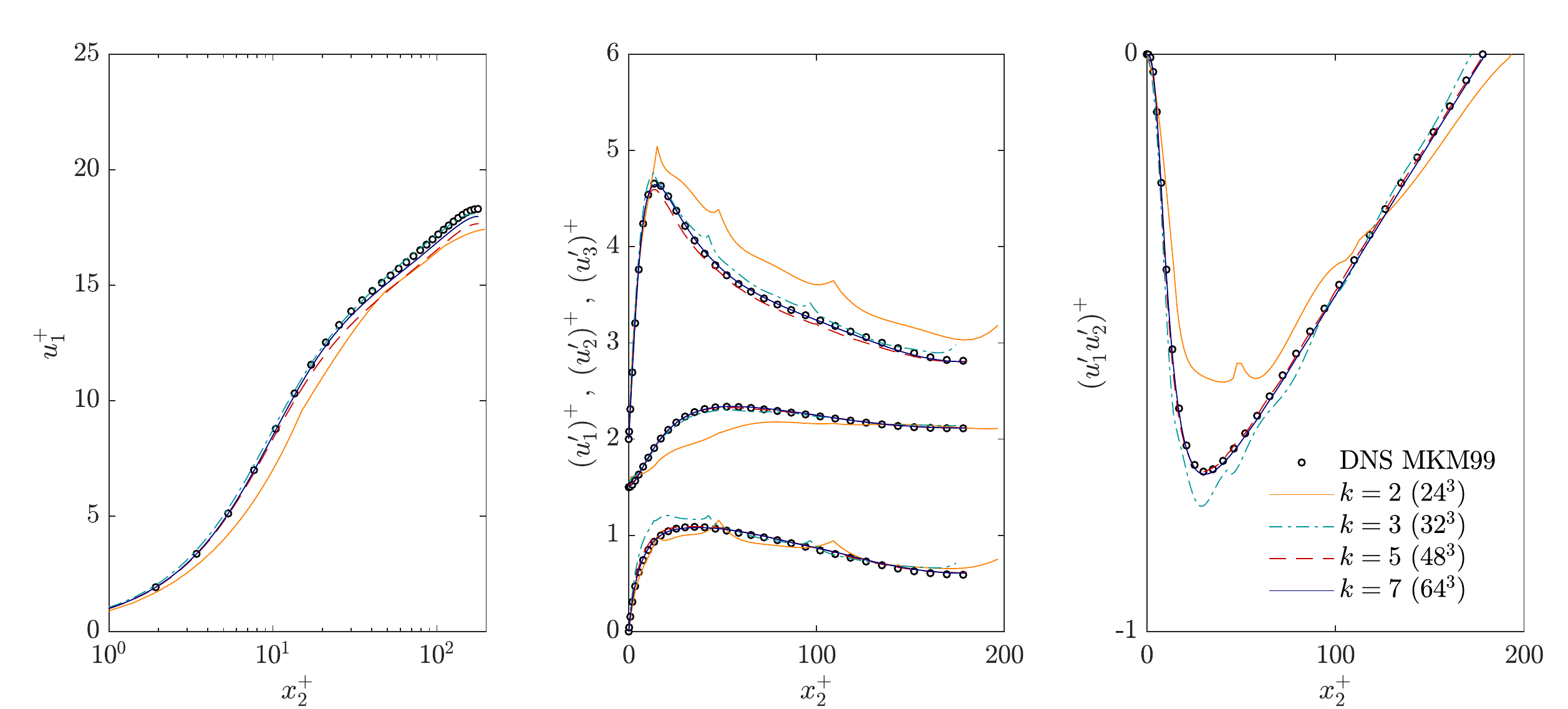}}
\caption{Turbulent channel flow at~$\mathrm{Re}_{\tau}=180$: Assessment of accuracy of results for polynomial degrees~$k=2,3,5,7$ with effective mesh resolutions of~$24^3$,~$32^3$,~$48^3$, and~$64^3$. For clarity, the profiles of~$\left(u_1^{\prime}\right)^+$ and~$\left(u_2^{\prime}\right)^+$ are shifted in vertical direction by~$2$ and~$1.5$, respectively.}
\label{fig:Turbulent_Channel_Re180}
\end{figure}

% HIGH-ORDER MAPPING
\begin{figure}[!ht]
 \centering 
 \subfigure[Compressible solver.]{
	\includegraphics[width=1.0\textwidth]{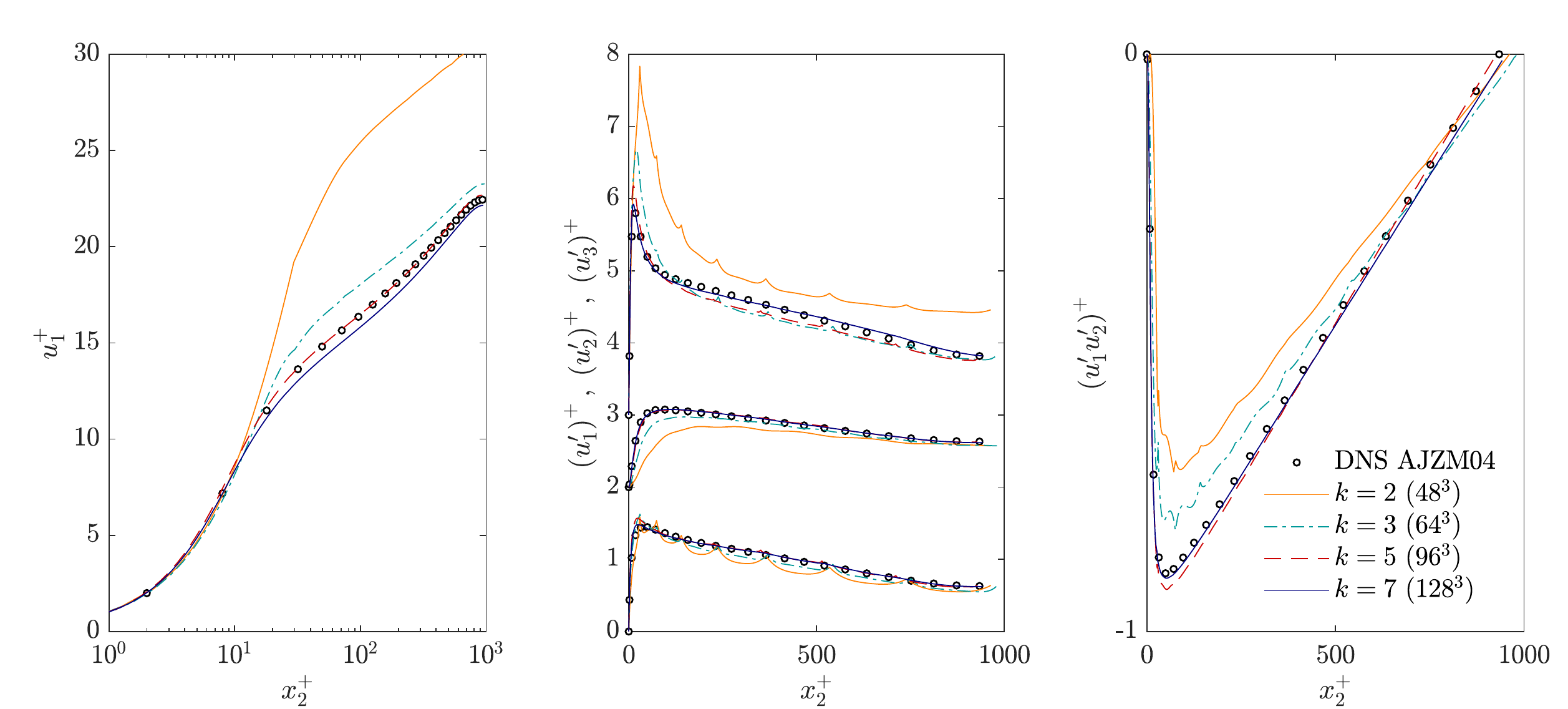}}
 \subfigure[Incompressible solver.]{
	\includegraphics[width=1.0\textwidth]{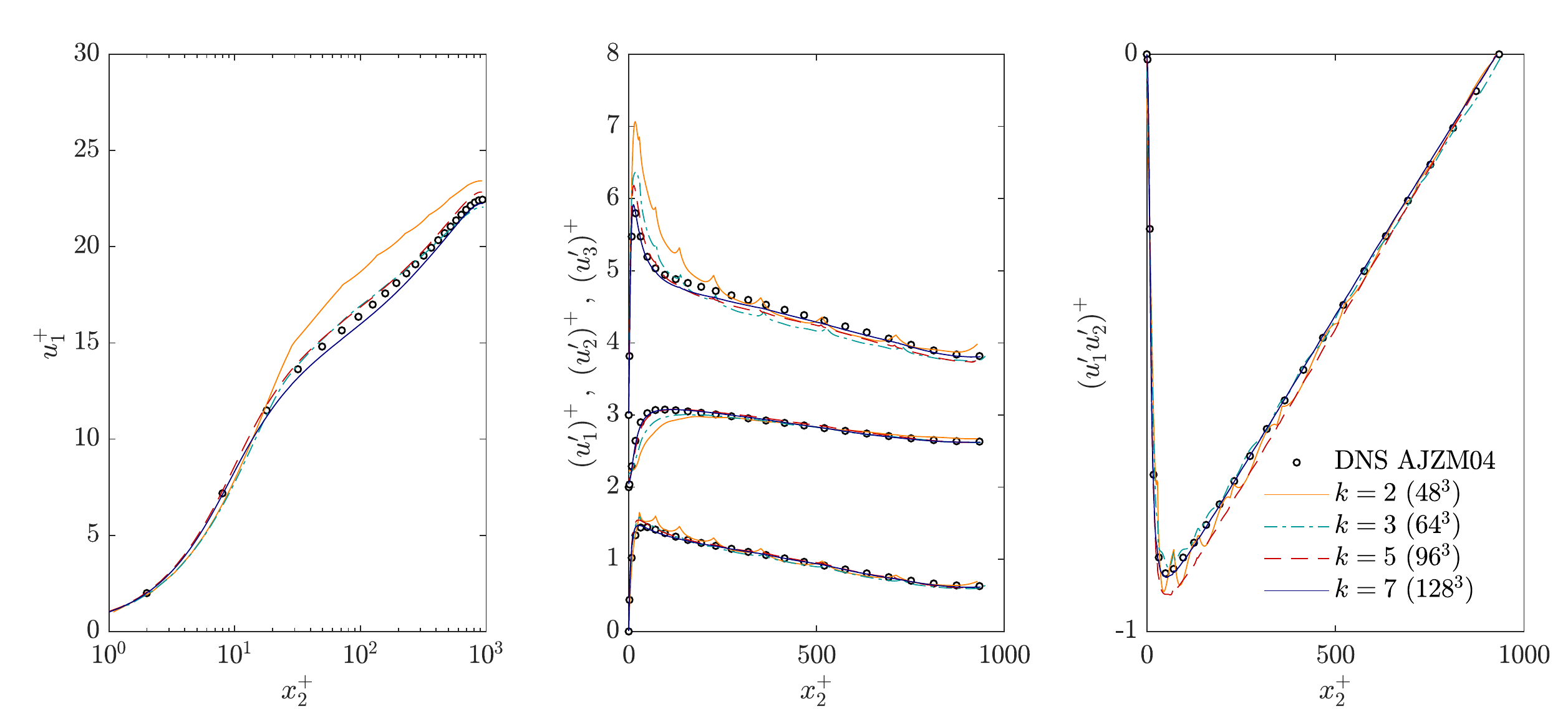}}
\caption{Turbulent channel flow at~$\mathrm{Re}_{\tau}=950$: Assessment of accuracy of results for polynomial degrees~$k=2,3,5,7$ with effective mesh resolutions of~$48^3$,~$64^3$,~$96^3$, and~$128^3$. For clarity, the profiles of~$\left(u_1^{\prime}\right)^+$ and~$\left(u_2^{\prime}\right)^+$ are shifted in vertical direction by~$3$ and~$2$, respectively.}
\label{fig:Turbulent_Channel_Re950}
\end{figure}

Figure~\ref{fig:Turbulent_Channel_Re180} shows results of statistical quantities for both the compressible and the incompressible solver at~$\mathrm{Re}_{\tau}=180$. The results converge towards the reference solution for increasing polynomial degree for both the compressible solver and the incompressible solver. While the results for the low-order method~$k=2$ deviate substantially from the reference solution, the profiles are captured very well for polynomial degrees~$k\geq 3$. Interestingly, the compressible DG solver appears to be more dissipative than the incompressible DG solver especially for~$k=2$. While the profile for the mean streamwise velocity is below the DNS reference data for the incompressible solver, it is above the reference solution for the compressible solver indicating a too dissipative method. The results for~$\mathrm{Re}_{\tau}=950$ are shown in Figure~\ref{fig:Turbulent_Channel_Re950}. A similar trend can be observed between the compressible formulation and the incompressible formulation for degree~$k=2$. Also for~$k=3$, the incompressible formulation is more accurate than the compressible one, which can be seen from the profiles of the mean streamwise velocity and the Reynolds shear stress. For the finer meshes with larger polynomial degrees~$k=(3,)5,7$ the results agree very well with the reference solution. We conclude that the incompressible solver tends to be more accurate for this flow problem but that both methods produce accurate results given the coarse resolutions and the fact that the turbulent flow solvers are parameter-free.

\subsubsection{Comparison of computational efficiency}\label{ComputationalCostsTurbulentChannel}
Finally, we investigate the computational costs for both solution strategies. The turbulent channel flow simulations have been performed on an Intel Haswell system (Intel Xeon CPU E5-2680 v3) for both the compressible solver and the incompressible solver. In Table~\ref{tab:Turbulent_Channel_Costs} we list the computational costs for the Reynolds numbers and spatial resolutions considered in the previous section. Since the overall computational costs depend linearly on the time interval~$t_{\mathrm{f}}$ and different time intervals are used in literature, the computational costs shown here are normalized to one flow-through time~$t_0$ and are specified without the costs for postprocessing and sampling of statistical results. The Courant number has been selected as large as possible for the different polynomial degrees in order to optimize the computational costs for the individual simulations. Accordingly, the potential gain in performance by using Courant numbers even closer to the critical Courant number can be expected to be very small and less than~$15\%$. To explain the observed differences in computational costs, we also list the ratio~$\Delta t_{\mathrm{inc}}/\Delta t_{s,\mathrm{comp}}$. As explained above, this ratio is significantly larger for the turbulent channel flow problem as compared to the Taylor--Green vortex example. In terms of overall computational costs, the incompressible solver is a factor of~$6.6-8.3$ faster for~$\mathrm{Re}_{\tau}=180$ and a factor of~$2.8-4.4$ for~$\mathrm{Re}_{\tau}=950$. The speed-up is smaller for~$\mathrm{Re}_{\tau}=950$ mainly due to the smaller Courant numbers in the range~$0.8-1.3$ required for the incompressible solver at~$\mathrm{Re}_{\tau}=950$ as compared to Courant numbers in the range~$1.3-1.8$ at~$\mathrm{Re}_{\tau}=180$. The results in Table~\ref{tab:Turbulent_Channel_Costs} reveal that solving one time step for the incompressible solver is approximately one order of magnitude more expensive than evaluating one Runge--Kutta stage for the compressible solver. All in all, the incompressible solver is significantly more efficient than the compressible solver for the turbulent channel flow problem characterized by stretched elements with large aspect ratios.

% HIGH-ORDER MAPPING, GRID_STRETCH_FACTOR = 1.8
% POSTPROCESSING: SAMPLING EVERY 10th TIME STEP, 11*k points per cell in y-direction for sampling
% COSTS WITHOUT POSTPROCESSING
\begin{table}
\caption{Turbulent channel flow problem: Comparison of computational efficiency of compressible high-order DG solver as compared to incompressible solver. The computational costs (without costs for postprocessing and sampling of statistical results) are specified in CPUh per flow-through time~$t_0$.}
\label{tab:Turbulent_Channel_Costs}
\renewcommand{\arraystretch}{1.1}
\begin{center}
\begin{tabular}{llllllllll}
\hline
& \multicolumn{4}{l}{$\mathrm{Re}_{\tau}=180$, refine level~$l=3$} & & \multicolumn{4}{l}{$\mathrm{Re}_{\tau}=950$, refine level~$l=4$}\\
 \cline{2-5} \cline{7-10}
& $k=2$ & $k=3$ & $k=5$ & $k=7$ & & $k=2$ & $k=3$ & $k=5$ & $k=7$\\
\hline
$\mathrm{Cr}_{\mathrm{comp}}$ 						 & 1.5 & 1.6 & 1.5 & 1.3 & & 1.6 & 1.8 & 1.8 & 1.6\\
$\mathrm{Cr}_{\mathrm{inc}}$  						 & 1.3 & 1.7 & 1.8 & 1.8 & & 1.3 & 1.1 & 0.8 & 0.9\\
$\Delta t_{\mathrm{inc}}/\Delta t_{s,\mathrm{comp}}$ & 67  & 82  & 92  & 107 & & 63  & 47  & 34  & 43\\
\hline
costs comp [CPUh/$t_0$] & $0.031$  & $0.099$ & $0.81$ & $3.7$  & & $0.66$ & $2.1$  & $17$  & $96$\\
costs inc [CPUh/$t_0$]  & $0.0042$ & $0.012$ & $0.11$ & $0.56$ & & $0.15$ & $0.53$ & $5.8$ & $34$\\
speed-up (inc vs.~comp) & $7.4$    & $8.3$   & $7.4$  & $6.6$  & & $4.4$  & $4.0$  & $2.9$ & $2.8$\\
\hline
\end{tabular}
\end{center}
\renewcommand{\arraystretch}{1}
\end{table}

\begin{remark}
We repeated the simulations for the incompressible solver using an adaptive time stepping technique with a local evaluation of the CFL condition according to~\cite{Krank2018} in order to estimate the possible performance improvement by using more advanced time stepping and Courant criteria. For~$\mathrm{Re}_{\tau}=180$, the gain in performance is marginal and not more than approximately~$5\%$. For~$\mathrm{Re}_{\tau}=950$, we could improve the overall computational costs by approximately~$20\%-50\%$ for the different polynomial degrees. This can be explained by the fact that the global Courant criterion used here has to yield a time step size that is stable for the whole simulation time, e.g., also for the initial transient from a prescribed initial condition to a developed turbulent flow, while the adaptive time stepping technique optimizes the time step size at every time~$t$. For the compressible solver, however, the gain in performance by applying adaptive time stepping can be expected to be even smaller than for the incompressible solver for small Mach numbers due to the more isotropic nature of the acoustic waves and the fact that the speed of sound (which is a function of the temperature) is approximately constant in space and time.
\end{remark}

\section{Summary}\label{Summary}
We have proposed a high-performance implementation for high-order discontinuous Galerkin discretizations of the compressible Navier--Stokes equations where the focus lies on the solution of turbulent incompressible flow problems in the present work. The solver is based on explicit time integration and a highly-efficient matrix-free implementation characterized by fast sum-factorization kernels and vectorization over several elements. We have analyzed the throughput of this implementation in terms of degrees of freedom processed per second demonstrating the high efficiency of our approach as compared to state-of-the-art implementations. The performance analysis reveals that the efficiency of the implementation is almost independent of the polynomial degree up to moderately high polynomial degrees suggesting that this approach is very efficient for discontinuous Galerkin spectral element simulations. We quantified the additional costs related to over-integration and showed that the corresponding performance penalty is significantly lower than expected theoretically. Accuracy and efficiency of this solver have been demonstrated for the 3D Taylor--Green vortex problem and the turbulent channel flow problem. To put the proposed approach into perspective, it has been compared to an incompressible Navier--Stokes solver using the same implementation framework. We introduced simple efficiency models to discuss the algorithmic differences between both formulations and the main ingredients impacting the performance of both approaches. We performed detailed and thorough numerical investigations to quantify the efficiency as well as the performance characteristics for practical computations. Our results indicate a clear performance advantage of several times for the incompressible formulation. This result is very interesting in the sense that mainly DG discretizations of the compressible Navier--Stokes equations have been used to solve incompressible turbulent flows so far.

\appendix

\section*{Acknowledgments}
The authors acknowledge discussions on compressible Navier--Stokes DG implementations with Christoph Haslinger. This research was partly funded by the German Research Foundation (DFG) under the project ``High-order discontinuous Galerkin for the EXA-scale'' (ExaDG) within the priority program ``Software for Exascale Computing'' (SPPEXA), grant agreement no. KR4661/2-1 and WA1521/18-1. The authors gratefully acknowledge the Gauss Centre for Supercomputing e.V.~(\texttt{www.gauss-centre.eu}) for funding this project by providing computing time on the GCS Supercomputer SuperMUC at Leibniz
Supercomputing Centre (LRZ, \texttt{www.lrz.de}) through project id pr83te.

%\section*{References}

\bibliography{paper}

\end{document}